\documentclass[twocolumn]{aastex631}

\usepackage{longtable}
\usepackage[flushleft]{threeparttable}
\usepackage{tabularx}
\usepackage{multirow}
\usepackage{graphicx}
\usepackage{amsmath,amssymb}
\usepackage{xcolor}
\usepackage{units}
\usepackage{epstopdf}
\usepackage{hyperref}
\usepackage{multirow}
\usepackage{url}
\usepackage{subfigure}
\usepackage{rotating}
\usepackage{gensymb}
\usepackage{enumitem}\setlist[description]{font=\textendash\enskip\scshape\bfseries}
\usepackage{booktabs}

\definecolor{navyblue}{rgb}{0.0, 0.0, 0.5}

\begin{document}

\title{Target of Opportunity Observations of Gravitational Wave Events \\with Vera C. Rubin Observatory}
\shorttitle{ToO Observations of GW Events with Vera C. Rubin Observatory}
\shortauthors{Andreoni et al.}

\author[0000-0002-8977-1498]{Igor Andreoni}
\altaffiliation{Gehrels Fellow}
\affil{Joint Space-Science Institute, University of Maryland, College Park, MD 20742, USA.}
\affil{Department of Astronomy, University of Maryland, College Park, MD 20742, USA.}
\affil{Astrophysics Science Division, NASA Goddard Space Flight Center, Mail Code 661, Greenbelt, MD 20771, USA.}
\email{andreoni@umd.edu}

\author[0000-0003-4768-7586]{Raffaella Margutti}
\affil{Department of Astronomy, University of California, Berkeley, CA 94720-3411, USA.}

\author[0000-0003-4924-7322]{{Om Sharan} Salafia}
\affil{INAF -- Osservatorio Astronomico di Brera, Via E. Bianchi 46, I-23807 Merate (LC), Italy.}
\affil{INFN -- Sezione di Milano-Bicocca, Piazza della Scienza 3, I-20146 Milano (MI), Italy.}

\author{B. Parazin}
\affil{College of Science, Northeastern University, Boston, Massachusetts 02115, USA.}

\author{V. Ashley Villar}
\affil{Department of Astronomy \& Astrophysics, The Pennsylvania State University, University Park, PA 16802, USA.}
\affil{Institute for Computational \& Data Sciences, The Pennsylvania State University, University Park, PA 16802, USA.}
\affil{Institute for Gravitation and the Cosmos, The Pennsylvania State University, University Park, PA 16802, USA.}

\author[0000-0002-8262-2924]{Michael W. Coughlin}
\affil{School of Physics and Astronomy, University of Minnesota, Minneapolis, Minnesota 55455, USA.}

\author{Peter Yoachim}
\affil{Department of Astronomy, University of Washington, 3910 15th Avenue NE, Seattle, WA 98195, USA}

\author[0000-0001-9676-5005]{Kris Mortensen}
\affiliation{Department of Physics, University of California, Davis, 1 Shields Avenue, Davis, CA 95616, USA.}

\author{Daniel Brethauer}
\affiliation{Department of Astronomy, University of California, Berkeley, CA 94720-3411, USA.}

\author{S. J. Smartt}
\affil{Astrophysics Research Centre
School of Mathematics and Physics,
Queen's University Belfast, Belfast, BT7 1NN, UK.}

\author{Mansi M. Kasliwal}
\affiliation{Division of Physics, Mathematics and Astronomy, California Institute of Technology, Pasadena, CA 91125, USA}

\author[0000-0002-8297-2473]{Kate D. Alexander}
\affil{Center for Interdisciplinary Exploration and Research in Astrophysics (CIERA) and Department of Physics and Astronomy, Northwestern University, 2145 Sheridan Road, Evanston, IL 60208-3112, USA.}

\author[0000-0003-3768-7515]{Shreya Anand}
\affil{Division of Physics, Mathematics and Astronomy, California Institute of Technology, Pasadena, CA 91125, USA}

\author[0000-0002-9392-9681]{E. Berger}
\affil{Center for Astrophysics -- Harvard \& Smithsonian, Cambridge, MA 02138, USA.}

\author{Maria Grazia Bernardini}
\affil{INAF -- Osservatorio Astronomico di Brera, Via E. Bianchi 46, I-23807 Merate (LC), Italy.}

\author[0000-0002-8576-1487]{Federica B. Bianco}
\affiliation{Department of Physics and Astronomy, University of Delaware, Newark, DE 19716, USA}
\affiliation{Joseph R. Biden, Jr.,  School of Public Policy and Administration, University of Delaware, Newark, DE 19717 USA}
\affiliation{Data Science Institute, University of Delaware, Newark, DE 19717 USA}
\affiliation{CUSP: Center for Urban Science and Progress, New York University, Brooklyn, NY 11201 USA}

\author{Peter K. Blanchard}
\affil{Center for Interdisciplinary Exploration and Research in Astrophysics (CIERA) and Department of Physics and Astronomy, Northwestern University, 2145 Sheridan Road, Evanston, IL 60208-3112, USA.}

\author{Joshua S. Bloom}
\affil{Department of Astronomy, University of California, Berkeley, CA 94720-3411, USA.}
\affil{Lawrence Berkeley National Laboratory, 1 Cyclotron Road, MS 50B-4206, Berkeley, CA 94720, USA.}

\author{Enzo Brocato}
\affil{INAF -- Osservatorio Astronomico di Roma, Via Frascati 33, I-00078 Monte Porzio Catone (RM), Italy.}
\affil{INAF -- Osservatorio Astronomico d’Abruzzo, Via M. Maggini s.n.c., I-64100 Teramo, Italy.}

\author[0000-0002-8255-5127]{Mattia Bulla}
\affil{The Oskar Klein Centre, Department of Astronomy, Stockholm University, AlbaNova, SE-10691, Stockholm, Sweden.}

\author{Regis Cartier}
\affil{Gemini Observatory, NSF's National Optical-Infrared Astronomy Research Laboratory, Casilla 603, La Serena, Chile.}

\author{S. Bradley Cenko}
\affil{Astrophysics Science Division, NASA Goddard Space Flight Center, Mail Code 661, Greenbelt, MD 20771, USA.}
\affil{Joint Space-Science Institute, University of Maryland, College Park, MD 20742, USA.}

\author[0000-0002-7706-5668]{Ryan Chornock}
\affiliation{Department of Astronomy, University of California, Berkeley, CA 94720-3411, USA.}

\author{Christopher M. Copperwheat}
\affil{Astrophysics Research Institute, Liverpool John Moores University, Liverpool, L3 5RF, UK.}

\author{Alessandra Corsi}
\affil{Texas Tech University, Lubbock, TX 79409, USA.}

\author{Filippo D'Ammando}
\affil{INAF -- Istituto di Radioastronomia, Via Gobetti 101, I-40129 Bologna, Italy.}

\author{Paolo D'Avanzo}
\affil{INAF -- Osservatorio Astronomico di Brera, Via E. Bianchi 46, I-23807 Merate (LC), Italy.}

\author{Laurence \'Elise H\'el\`ene Datrier}
\affil{Institute for Gravitational Research, University of Glasgow, Glasgow, UK}

\author{Ryan J. Foley}
\affil{Department of Astronomy and Astrophysics, University of California, Santa Cruz, CA 95064, USA.}

\author{Giancarlo Ghirlanda}
\affil{INAF -- Osservatorio Astronomico di Brera, Via E. Bianchi 46, I-23807 Merate (LC), Italy.}

\author{Ariel Goobar}
\affil{The Oskar Klein Centre, Department of Physics, Stockholm University, AlbaNova, SE-106 91 Stockholm, Sweden.}

\author{Jonathan Grindlay}
\affil{Center for Astrophysics -- Harvard \& Smithsonian, Cambridge, MA 02138, USA.}

\author{Aprajita Hajela}
\affil{Center for Interdisciplinary Exploration and Research in Astrophysics (CIERA) and Department of Physics and Astronomy, Northwestern University, 2145 Sheridan Road, Evanston, IL 60208-3112, USA.}

\author{Daniel E. Holz}
\affil{Department of Physics, Department of Astronomy \& Astrophysics, Kavli Institute for Cosmological Physics, and Enrico Fermi Institute, The University of Chicago, Chicago, IL 60637, USA.}

\author{Viraj Karambelkar}
\affil{Division of Physics, Mathematics and Astronomy, California Institute of Technology, Pasadena, CA 91125, USA}

\author{E. C. Kool}
\affil{The Oskar Klein Centre, Department of Astronomy, Stockholm University, AlbaNova, SE-10691, Stockholm, Sweden.}

\author{Gavin P. Lamb}
\affil{School of Physics and Astronomy, University of Leicester, University Road, Leicester, LE1 7RH, UK.}

\author[0000-0003-1792-2338]{Tanmoy Laskar}
\affiliation{Department of Astrophysics/IMAPP, Radboud University Nĳmegen, P.O. Box 9010, 6500 GL Nijmegen, The Netherlands.}

\author{Andrew Levan} 
\affil{Department of Astrophysics/IMAPP, Radboud University, PO Box 9010, 6500 GL, The Netherlands.}
\affil{Department of Physics, University of Warwick, Coventry, CV4 7AL, UK.}

\author[0000-0002-9770-3508]{Kate Maguire}
\affil{School of Physics, Trinity College Dublin, The University of Dublin, Dublin 2, Ireland.}

\author{Morgan May}
\affil{Department of Physics, Columbia University, New York, NY 10027, USA.}

\author{Andrea Melandri}
\affil{INAF -- Osservatorio Astronomico di Brera, Via E. Bianchi 46, I-23807 Merate (LC), Italy.}

\author[0000-0002-0763-3885]{Dan Milisavljevic}
\affiliation{Department of Physics and Astronomy, Purdue University, 525 Northwestern Avenue, West Lafayette, IN 47907, USA.}

\author[0000-0001-9515-478X]{A.~A.~Miller}
\affiliation{Center for Interdisciplinary Exploration and Research in Astrophysics (CIERA) and Department of Physics and Astronomy, Northwestern University, 2145 Sheridan Road, Evanston, IL 60208-3112, USA.}

\author{Matt Nicholl}
\affil{Institute for Gravitational Wave Astronomy and School of Physics and Astronomy,\\ University of Birmingham, Birmingham B15 2TT, UK.}

\author{Samaya M. Nissanke}
\affil{GRAPPA, University of Amsterdam, Science Park 904, 1098 XH Amsterdam, Netherlands}

\author{Antonella Palmese}
\altaffiliation{NASA Einstein Fellow}
\affiliation{Department of Physics, University of California Berkeley, 366 LeConte Hall MC 7300, Berkeley, CA, 94720, USA}

\author{Silvia Piranomonte}
\affil{INAF -- Osservatorio Astronomico di Roma, Via Frascati 33, I-00078 Monte Porzio Catone (RM), Italy.}

\author{Armin Rest}
\affil{Space Telescope Science Institute, 3700 San Martin Drive, Baltimore, MD 21218, USA.}
\affil{Department of Physics and Astronomy, The Johns Hopkins University, 3400 North Charles Street,
Baltimore, MD 21218, USA.}

\author{Ana Sagu\'es-Carracedo}
\affil{The Oskar Klein Centre, Department of Physics, Stockholm University, AlbaNova, SE-106 91 Stockholm, Sweden.}

\author{Karelle Siellez}
\affil{Institut d’Astrophysique de Paris, CNRS, UMR 7095, 98 bis bd Arago, 75014 Paris, France.}

\author{Leo P. Singer}
\affil{Astroparticle Physics Laboratory, NASA Goddard Space Flight Center, Mail Code 661, Greenbelt, MD 20771, USA.}

\author{Mathew Smith}
\affil{Universit\'e de Lyon, Universit\'e Claude Bernard Lyon 1, CNRS/IN2P3, IP2I Lyon, F-69622, Villeurbanne, France.}

\author[0000-0003-0771-4746]{D. Steeghs}
\affil{Department of Physics, University of Warwick, Gibbet Hill Road, Coventry CV4 7AL, UK.}
\affil{OzGrav: The ARC Centre of Excellence for Gravitational Wave Discovery, Clayton VIC 3800, Australia.}

\author{Nial Tanvir}
\affil{School of Physics and Astronomy, University of Leicester, University Road, Leicester, LE1 7RH, UK.}

\begin{abstract}
The discovery of the electromagnetic counterpart to the binary neutron star merger GW170817 has opened the era of gravitational-wave multi-messenger astronomy. Rapid identification of the optical/infrared kilonova enabled a precise localization of the source, which paved the way to deep multi-wavelength follow-up and its myriad of related science results. 
Fully exploiting this new territory of exploration requires the acquisition of electromagnetic data from samples of neutron star mergers and other gravitational wave sources. After GW170817, the frontier is now to map the diversity of kilonova properties and provide more stringent constraints on the Hubble constant, and enable new tests of fundamental physics. The Vera C. Rubin Observatory's Legacy Survey of Space and Time (LSST) can play a key role in this field in the 2020s, when an improved network of gravitational-wave detectors is expected to reach a sensitivity  that will enable the discovery of a high rate of merger events involving neutron stars ($\sim$tens per year) out to distances of several hundred Mpc. 
We design comprehensive target-of-opportunity observing strategies for follow-up of gravitational-wave triggers that will make the Rubin Observatory the premier instrument for discovery and early characterization of neutron star and other compact object mergers, and yet unknown classes of gravitational wave events.

\end{abstract}

\section{Introduction}
\label{sec:intro}

The direct detection of gravitational waves (GW) from astrophysical sources has enabled an exciting new view of the cosmos \citep{GWdiscovery}. The true power of GW detections becomes apparent when they are paired with  electromagnetic (EM) data.

To date, the first and only celestial object with confirmed joint GW+EM detections was GW170817, which was discovered in association with a short gamma-ray burst \citep[GRB;][]{Abbott2017gw_grb, Goldstein2017}, an optical kilonova \citep[KN; e.g.][]{Coulter17,Valenti:2017ngx,Arcavi2017GW,Tanvir17,Lipunov2017,Soares-Santos2017}, and a radio \citep[e.g.,][]{Alexander2017, Hallinan:2017woc} and X-ray \citep[e.g.,][]{Troja2017, Margutti:2017cjl} afterglow.
The identification of an EM counterpart provides numerous benefits to GW analysis, including: improved localization leading to host-galaxy identification \citep[e.g.,][]{Coulter17}; determination of the source's distance and energy scales; characterization of the progenitor's local environment \citep[e.g.,][]{Alexander2017, Hallinan:2017woc, Levan2017, Pan2017, Troja2017, Hajela2019}; breaking modeling degeneracies between distance and inclination \citep{NSNSrate}; insight on the launching and propagation of relativistic jets, and the related emission processes \citep[e.g.,][]{Murguia-Berthier2014, Gottlieb2018, Mooley2018Nat, Ghirlanda2019Sci, Nativi2021a, Nativi2021b, Salafia2019, Salafia2021}; and insight on the formation channel of binary neutron star mergers \citep[e.g.,][]{Palmese2017}. Furthermore, identification of the EM counterpart facilitates other fields of investigation such as determining the primary sites of heavy, rapid neutron capture ``$r$-process" element production \citep{ChBe2017,Coulter17, Cowperthwaite17, Kasen17,Kilpatrick2017, PiDa2017,RoFe2017,Smartt17,Rosswog2018, WaHa2019,KaKa2019}, placing limits on the neutron star (NS) equation of state \citep{BaJu2017, MaMe2017, CoDi2018, CoDi2018b, CoDi2019b, AnEe2018, MoWe2018,RaPe2018,Lai2019,DiCo2020, Nicholl2021}, and making independent measurements of the Hubble constant \citep{AbbottNSdiscovery,AbbottH0, Guidorzi2017, Hjorth17, HoNa2018, CoDi2019,CoAn2020, DiCo2020, Wang2021H0}. We refer the reader to \cite{Nakar2020PhR} and \cite{Margutti21} for recent reviews of the GW and EM observations of GW170817. 

The third Advanced LIGO, Virgo, and KAGRA (LVK) observing run (O3, which ran in 2019--2020) yielded the solid detection of the second binary neutron star (NS--NS) merger \citep[GW190425;][]{Abbott:2020uma}, at least two neutron star--black hole (NS--BH) mergers \citep[GW200105 and GW200115;][]{Abbott2021NSBH}, and several other NS--NS or NS--BH candidates \citep{LIGO_GWTC2_pop_2020}. Despite much follow-up effort, no EM counterpart was identified during O3 in the optical \citep[e.g.,][]{Andreoni2019S190510g, CoAh2019b, Goldstein2019S190426c, Gomez2019, Hosseinzadeh2019, LuPa2019, Ackley2020, Andreoni2020S190814bv, AnAg2020, Garcia2020, Gompertz:2020cur,Kasliwal2020, Vieira2020, AnCoNatas2021, Chang2021, Kilpatrick2021arXiv, Oates2021, Becerra2021MNRAS}, in the radio \citep{Dobie2019, Alexander2021arXiv, Bhakta2021}, or during X--ray/high-energy observations \citep{Page2020,  Watson2020} \citep[see however][]{Pozanenko2020}.  The task was made particularly difficult by the coarse localization regions \citep[median localization area of 4480\,deg$^2$;][]{Kasliwal2020} and large distances \citep[median distance of 267\,Mpc;][]{Kasliwal2020} of NS--NS and NS--BH merger candidates \citep[see also][]{GWTC-2.1arXiv}.

Exploiting the success of multi-messenger astronomy in the next decade will require a continued investment of observational resources. In this period, the GW detector network will increase its sensitivity, while additional interferometers will come online, such as LIGO-India \citep{Abbott2020LRRprospects}. In this multi-detector regime, NS--NS mergers will be detected beyond $\sim 200$~Mpc and NS--BH mergers out to several hundred Mpc. Nearby source localizations will continue to improve from $\sim$$100$~deg$^2$ to $\sim$$10$~deg$^2$ for those mergers detected by multiple interferometers \citep{PeSi2021}.
Vera C. Rubin Observatory will have a unique combination of large aperture and wide field-of-view that will be well suited to the task of GW follow-up. Moreover, LSST will provide deep multi-band templates of $>18,000$ deg$^2$ for immediate image subtraction, which is key to transient discovery.  Rubin will be able to cover well-localized GW regions in a handful of pointings and achieve deep observations with relatively short integration times. This means that Rubin has the potential to detect and identify EM counterparts to GW sources rapidly and effectively, especially at such large distances, where counterparts (M$\sim -16$\,mag in the optical) are expected to be too faint for most wide-field survey telescopes \citep[e.g.,][]{Bloom2009arXiv, Chase2021arXiv}. However, rapid target of opportunity (ToO) observations will be the only way to achieve this goal.

In this paper, which is largely based on the white paper by \cite{Margutti2018WP}, we describe comprehensive ToO strategies for the follow-up of GW sources that will allow Rubin to serve as the premiere discovery instrument in the southern hemisphere. The start of science operations of Rubin is set in 2024+, hence it will overlap with the fifth LIGO-Virgo-KAGRA observing run (LVK O5).
The fourth LVK observing period (O4) will run mid 2022--23 and, with the increased sensitivity from O3, is projected to discover up to tens of NS--NS mergers (Table\,\ref{tab:detection_prospects}). However, this is an optimistic estimate with large uncertainty and 40--50\% will likely be in solar conjunction, thus by O5 one can only expect incremental increase in EM counterpart discovery. Rubin will be the next game changer.

We outline two LSST observing strategies based on  the expected performance of GW detectors during O5: a {\it minimal} strategy that targets a time investment of $\lesssim 1.4\%$ of the nominal survey time and an {\it preferred} strategy that will use $\sim 2\%$ of the time budget. These strategies are designed to provide rapid discovery of EM counterparts, which will enable further multi-wavelength photometric and spectroscopic observations. Our work tackles the following major science goals: 

\textbf{[i]} The primary goal that will enable studies of EM transients from GW sources in the 2020s is growing the sample size of known EM counterparts.

Building a large sample of EM counterparts is essential for conducting statistically rigorous systematic studies that will allow us to understand the diversity of EM transient behavior, their host environments, the nature of merger remnants, and their contribution to the chemical enrichment of the universe through cosmic $r$-process production, which shapes the light-curves and colors of KNe associated to GW events \citep[e.g.,][]{2015MNRAS.446.1115M}. In fact, the KN population is {\it expected} to be diverse, since simulations suggest that the ejected masses and lanthanide fractions (hence observable properties such as color, luminosity, and spectral features) are significantly dependent on the binary mass ratio \citep[see for example][for a recent review]{Radice2020ARNPS}.

Improvements in survey data mining technology will enable the discovery of rare KNe in the Wide Fast Deep (WFD) survey \citep{Cowperthwaite18, Scolnic2018, Andreoni2019LSST, Bianco2019, Setzer2019MNRAS, Andreoni2021RubinarXiv, Sagues2021}.
However, targeted follow-up will be much more efficient at achieving this goal thanks to timing and search-area constraints provided by GW detections. The chances of detecting a KN associated with a GW event during the regular WFD survey, without initiating ToO observations, is negligible \citep[although ``reverse" searches for faint signals in GW data that could be associated with EM-discovered transients is an intriguing prospect, see for example][]{Aasi2013PhRvD..88l2004A}. Moreover, a multi-messenger dataset (as opposed to EM-only studies) carries much higher scientific value \citep[e.g.,][]{DiCo2020}. 

\textbf{[ii]} Of particular interest are observations of KNe at early times (e.g., $\lesssim 11$~hr post-merger). Despite the fact that the optical counterpart of GW170817 was discovered 10.9\,hr post-merger \citep{Coulter17} \citep[see also e.g.][]{Andreoni2017GW, Arcavi2017GW,Cowperthwaite17,Drout17, Evans2017, Kasliwal17,Lipunov2017, Pian2017, Smartt17,Soares-Santos2017,Tanvir17,Valenti17,Villar17}, these observations were still unable to definitively determine the nature of the early blue emission. Understanding this early-time radiation is crucial for identifying emission mechanisms beyond the radioactively powered KN \citep[such as a precursor from $\beta$ decay of free neutrons, or shock-cooling, see for example][]{2015MNRAS.446.1115M, Arcavi2018, PiroKollmeier18}. In particular, mapping the rapid broad-band spectral energy distribution (SED) evolution is key to separating these components, and also distinguishing KN from most other astrophysical transients. Photometric observations in multiple bands can serve well for this purpose. If a bright ($\lesssim 21.5$ mag) counterpart is identified rapidly enough, precious spectroscopic data can be acquired that offer an even better opportunity of differentiating between those mechanisms.

\textbf{[iii]} An EM counterpart to a NS--BH merger is yet to be observed \citep[e.g.,][]{AnCo2020}. In this case, the merger might produce a KN \citep[e.g.,][]{Li1998, Roberts:2011xz, Foucart2012, Kawaguchi2016, Barbieri2020}, but the ejecta mass can vary significantly (from $\sim$zero to $\sim 0.5$\,M$_{\odot}$) depending on the mass ratio of the binary, the NS equation of state, and the BH spin \citep[e.g.][]{Foucart:2012vn, Kawaguchi2016, Gompertz2021arXiv}. 
It is also unclear if NS--BH mergers will be able to produce the bright early-time blue emission seen in GW170817 \citep{2015MNRAS.446.1115M}, if any EM transient is produced at all. Furthermore, these systems will have higher amplitude GWs and will thus be detected on average at greater distances, as O3 demonstrated \citep{LIGO_GWTC2_pop_2020, Abbott2021NSBH}. This combination of increased luminosity distance and potentially fainter counterpart means that Rubin will be an essential tool for discovering (or placing the deepest limits on) their EM counterparts.

\textbf{[iv]} Rubin, equipped with ToO capabilities, has the potential to place deep limits on the optical emission from binary black hole (BH--BH) mergers. There are numerous speculative mechanisms for the production of an optical counterpart to a BH--BH merger \citep[e.g.,][]{Perna,Loeb,Stone,deMink,McKernan},
yet none has been unambiguously observed. One candidate optical flare, which might be associated to the BH--BH merger GW190521, was found by \cite{Graham2020PhRv}. Rubin will be able to place deep limits on the optical emission from BH--BH mergers with a high statistical confidence in the case of non-detections, or might be able to discover the first high confidence EM counterpart to BH--BH mergers.

\textbf{[v]} Lastly, Rubin has the capabilities to explore the currently uncharted territory of EM counterparts to yet-to-be identified GW sources which are of burst nature and not modelled by compact object coalescence \citep[e.g., from a nearby core-collapse SN, cf.][]{Kotake2006RPPh}. 

\vskip +0.1 cm

In the pursuit of these goals, the true power of Rubin will be the ability to both rapidly grow the population of rare known transients, such as KNe, and discover new sources of optical emission associated with compact object mergers (e.g., non-radioactively powered KN early-time emission, emission from a BH--BH merger) and unidentified GW sources.

%%%%%%%%%%%%%%%%%%%%%%%%%%%%%%%%%%%%%%%%%%%%%%%%%%%%%%%%%%%%%%%%%%%%%%%%%%%%%%%%%%
\section{Technical Description}
\label{sec: technical}

%------------------------------------------------------------------------------
\subsection{High-level description}
\label{subsec: high-level-description}

The likelihood that, during the LSST WFD survey, the coordinates of a counterpart fall within the Rubin field of view (FoV) by chance multiple times within $\sim1$ week since a GW trigger was found to be extremely small \citep[$\sim7\%$ for  $r$-band only; $\sim$ a few $\%$ for observations in multiple filters;][]{Margutti2018WP}. This conclusion has been significantly strengthened by studies that focused on the problem of the detection and characterization of KNe from NS--NS mergers in the WFD data stream using realistic simulations of the observing cadence and conditions. These studies either started from re-scaled versions of  the single known KN event with multi-band light-curves \citep{scolnic2017, Bianco2019}, or expanded this specific case with simulations of KN light-curves expected for a wide range of ejecta masses and composition \citep{Cowperthwaite18, Andreoni2019PASP, Setzer2019MNRAS, Andreoni2021RubinarXiv, Sagues2021}, and viewing angles \citep{Andreoni2021RubinarXiv, Sagues2021}. 

The main findings from these studies can be summarized as follows: (i) The main LSST survey will reach an overall efficiency of KN detection\footnote{The definition of what constitutes a detection varies from study to study, but generically implies the capability to detect with high statistical confidence the KN emission in one or multiple bands and in at least one instance in time, and reject asteroids.} of the order of a few $\%$. For the optimistic end of the NS--NS merger rate $R_{\text{BNS}} = 320^{+490}_{-240}$\,Gpc$^{-3}$yr$^{-1}$ \citep{LIGO_GWTC2_pop_2020}, results from these works (with neutron star merger rates appropriately re-scaled) generally agree that 1--4 GW170817-like KNe  per year will be detected in the LSST WFD using the baseline cadence, and $\sim0.3$ KNe per year in the LSST Deep-Drilling Fields (DDFs). 
 (ii) While the optimistic prospect of finding up to 4 KNe per year might seem encouraging, the vast majority of the detected KNe will have poorly sampled light-curves, which can prevent accurate
 estimates of physical parameters of  primary scientific importance such as the merger ejecta mass and electron fraction $Y_e$. KNe discovered this way will also likely be found $>24$\,hr from the merger, which will prevent the study of the possible fast-fading blue component. One major challenge will be effectively separating those handful of KNe from contaminant sources, whose number can be several orders of magnitude larger \citep[but see for example][for techniques to make this separation more effective]{Andreoni2021RubinarXiv, Andreoni2021ztfrest}. In addition, those KN detections will lack any GW information that could give insight in the determination of the progenitor and the physics of the merger. 
 
 These two results are direct consequences of the fact that the cadence of the LSST WFD survey is inadequate given the expected fast evolution of GW counterparts \citep[see also][]{Bellm2021arxiv}, and that the sky area covered by the DDFs is not large enough to rely on chance alignment with GW localizations. Further improvement on the LSST WFD survey design with implementation of rolling cadences could lead to the discovery of a significantly larger number of KNe independently of GW or GRB triggers, which is key to unbiased studies of the KN population beyond the LVK horizon and from all viewing angles. Nevertheless, ToO capabilities are the only way to enable Rubin to have a significant scientific role in joint GW+EM multi-messenger Astrophysics, for NS--NS as well as NS--BH and BH--BH mergers. As demonstrated below, only a small amount of LSST survey time during the O5 run is required in order to make a major scientific contribution. 

In this section, we analyze separately the cases of ToO follow-up of GW triggers resulting from NS--NS mergers, NS--BH mergers, BH--BH mergers as well as un-modeled GW sources. For each of these classes we outline a {\it minimal} and  {\it preferred} Rubin follow-up strategy. We design the follow-up strategies of GW triggers bearing in mind that at the time of writing we have only one example of well observed KN from the NS--NS merger event GW170817 (unambiguous EM counterparts to NS--BH and BH--BH mergers are yet to be found), and that our knowledge of EM counterparts to GW events could improve in the next few years \emph{before} the start of Rubin operations. The strategies that we are putting forward see sudden changes when the localization area passes a given threshold (for example 20\,deg$^2$). In reality, uncertainty in the localization area measurement should be taken into account and a more conservative choice for the integrated probability contour (e.g., 95\% instead of 90\%) could be considered for exceptionally promising GW events. We propose that these strategies are used as robust guidelines, with some flexibility allowed at the time of their application.

In the 2024+ era of LSST operations, the sky localization regions from a four GW-detector network operating at design sensitivity will routinely \citep[but not typically][]{PeSi2021} be of the order of  20--200 deg$^2$, depending on distance, sky location and orientation of the merger event \citep{Abbott2020LRRprospects}.
Although the impact (and timeline) of KAGRA and LIGO-India are still uncertain, areas of tens of deg$^2$ may become common and time windows with at least three online detectors will increase, improving the overall distribution of sky localizations for detections.
Rubin has a unique combination of capabilities for optical/near-IR counterpart searches: the $\sim$10 deg$^2$ camera, deep sensitivity (over 6 bands) and a deep sky template for subtraction after the first year of operations.
In addition, the fast readout and slew times are ideally suited to fast mapping of $20$--$200$\,deg$^{2}$ areas, which are not expected to be typical but can become routine during O5 \citep{PeSi2021}, to depths that are untouchable by the other surveys currently in this search and discovery mission. 

Facilities such as the Asteroid Terrestrial-impact Last Alert System (ATLAS; \citealt{ToDe2018}), the Zwicky Transient Facility (ZTF; \citealt{Bellm:19:ZTFScheduler,Graham2019PASP}) and Gravitational-wave Optical Transient Observer \citep[GOTO;][]{Steeghs2021arXiv} can cover large areas with their cameras, but do not have the aperture to go beyond magnitude $21$--$22$ and have limited filter sets. The Panoramic Survey Telescope and Rapid Response System \citep[Pan-STARRS;][in the Northern hemisphere]{ChMa2016} and the Dark Energy Camera \citep[DECam;][in the Southern hemisphere]{FlDi2015} are mounted on larger telescopes and therefore more sensitive. Compared to DECam, Rubin has the following key advantages: a larger FoV (9.6 deg$^2$ against $\sim 3$\,deg$^2$ of DECam), larger collecting area (which makes Rubin significantly more sensitive), shorter read-out time and the advantage of having an all-sky reference frame with which to do immediate transient discovery via image subtraction. Other planned facilities include BlackGEM \citep{BlGr2015}, a southern hemisphere GOTO node, and the La Silla Schmidt Southern Survey (LS4), which are also limited in aperture and sensitivity compared to Rubin.

Rubin is expected to start operations in 2024.
Comparing the timelines of the Rubin and the GW observatories projects, Rubin will become operational by the start of the fifth observing run (O5).  

For the observability of individual GW events, we assume that Rubin can access roughly 2/3 of the sky, which is generous because follow-up might be performed only for events falling within the LSST footprint ($\sim 18,000$\,deg$^2$), or where templates are available in at least one band. To ensure that GW localization skymaps are properly covered, we consider $\sim \times 2$ the minimum number of pointings when developing the strategies, So that chip gaps can be covered by applying small offsets between consecutive observations. For instance, a sky area of 20\,deg$^2$ could be imaged with two Rubin pointings ($\sim 10$\,deg$^2$ each), but four tiles are considered instead to avoid losing $\sim 4\%$ of the area in any band due to gaps between the detectors.  We apply a usable weather correction of 80\% based on Cerro Tololo historical records.

The LSST camera is equipped with 5 filter slots. This means that observations with all six $u$+$g$+$r$+$i$+$z$+$y$ filters will not be possible to obtain in a given night. In dark nights, the $u$ filter will be available, but the $z$ filter will not. Conversely, $z$ filter will be available in bright nights, but the $u$ filter will not. Therefore  the exact GW follow-up strategies will be slightly different depending on the moon phase. For time budget calculations, we assume 7\,s of overhead time between exposures and 120\,s overhead time for each filter change.

%---------------------------------------------------------
\subsubsection{Binary Neutron Star mergers (NS--NS)}
\label{SubSubSec:NSNS}

For NS--NS mergers we identify two key areas of the parameter space that can be explored by Rubin better than any other existing optical telescope: (i) the very early ($\delta t<12$\,hr) multi-band evolution of the KN emission; (ii) the faint end of the KN brightness distribution. We expect numerous faint KNe resulting from distant mergers or from intrinsically low-luminosity events that populate the faint end of the KN luminosity function. We design the Rubin follow-up strategy of NS--NS mergers around the two discovery areas above.
By sampling the rise time of the KN emission in multiple bands, the Rubin will enable constraints on new emission components such as shock cooling emission (proposed for GW170817 by \citealt{PiroKollmeier18}) or a free neutron precursor \citep{2015MNRAS.446.1115M}. 

Other survey instruments in the Southern hemisphere do not reach a comparable depth and, because of their smaller FoV, will have to tile the GW localization region with several pointings. The combination of those two factors -- large 10\,deg$^2$ FoV and unique depth -- make Rubin a particularly efficient at early KN discovery. The multi-band exploration of the very early KN emission is a key strength of the Rubin GW follow-up program that we propose here.

\begin{figure*}
\center{\includegraphics[scale=0.275]{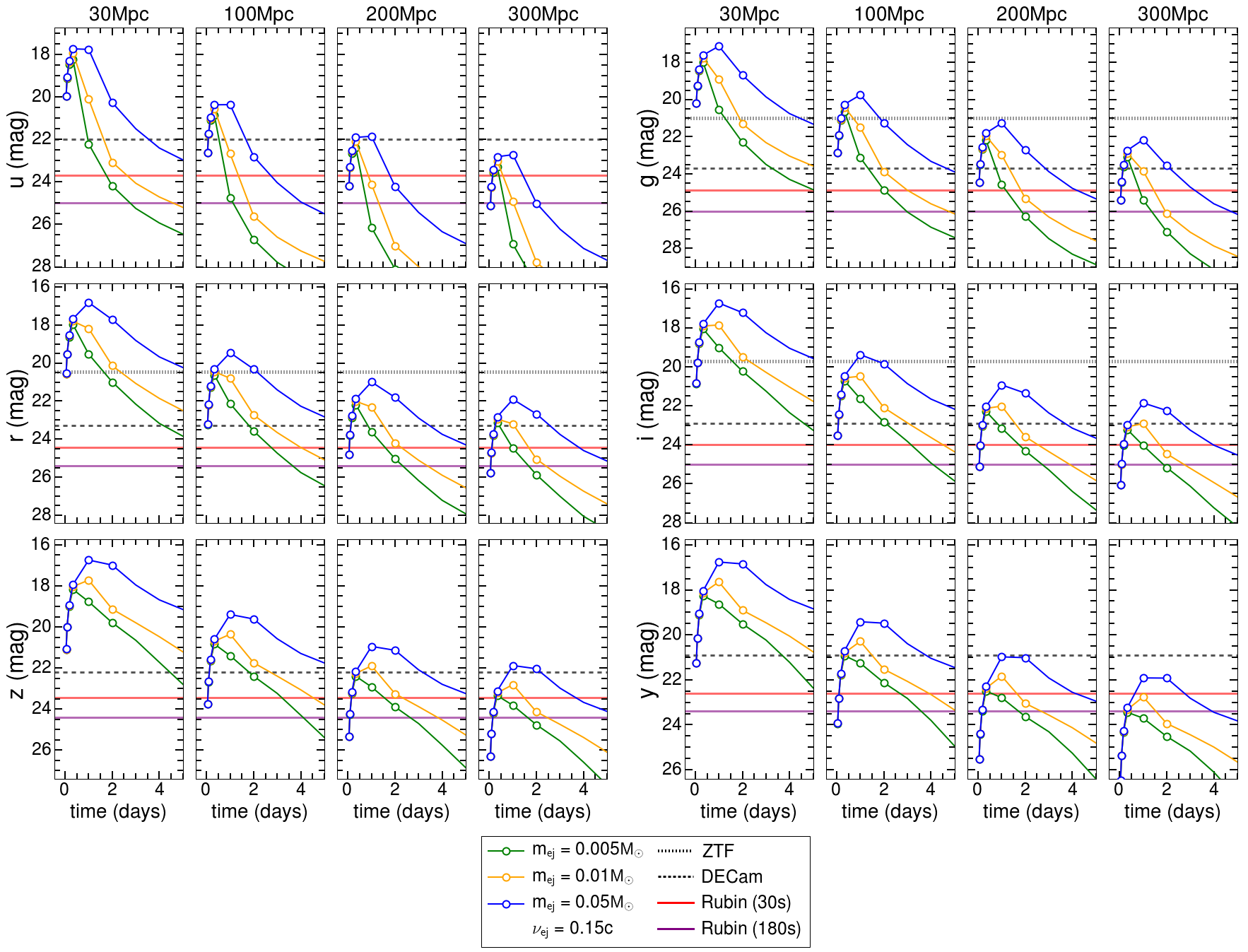}}
\caption{Simulated KN light-curves in the six Rubin filters for different properties of the ejecta (mass and velocity) at four representative distances (30, 100, 200 and 300 Mpc). The models include a ``red''  and ``blue'' KN component. We explore three values of the red-component ejecta  mass, $M_\mathrm{ej,R}=0.005,0.01, 0.05\,\rm{M_{\odot}}$, with velocity $v_\mathrm{ej,R}=0.15\,c$ (the KN luminosity is not a strong function of $v_\mathrm{ej,R}$ and values within $0.1$--$0.2\,c$ give comparable results). For each combination of these parameters, the blue ejecta component is $M_\mathrm{ej,B}= 0.5
\times M_\mathrm{ej,R}$ and $v_\mathrm{ej,B}= 1.5\times v_\mathrm{ej,R}$. Open circles depict the expected preferred cadence times post merger (1, 2, 4, 24, and 48\,hr, with the possible addition of data at 8\,hr). Dotted and dot-dashed horizontal lines mark typical $5\sigma$ detection thresholds of ZTF and DECam, respectively, assuming 30\,s exposure times (although GW follow-up with those instruments is likely to be performed using longer exposure times). Red and purple solid lines: Rubin $5\sigma$ detection thresholds for exposure times of 30\,s and 180\,s under ideal observing conditions.}
 \label{Fig:KNredblue}
\end{figure*}

\begin{figure*}
\center{\includegraphics[scale=0.275]{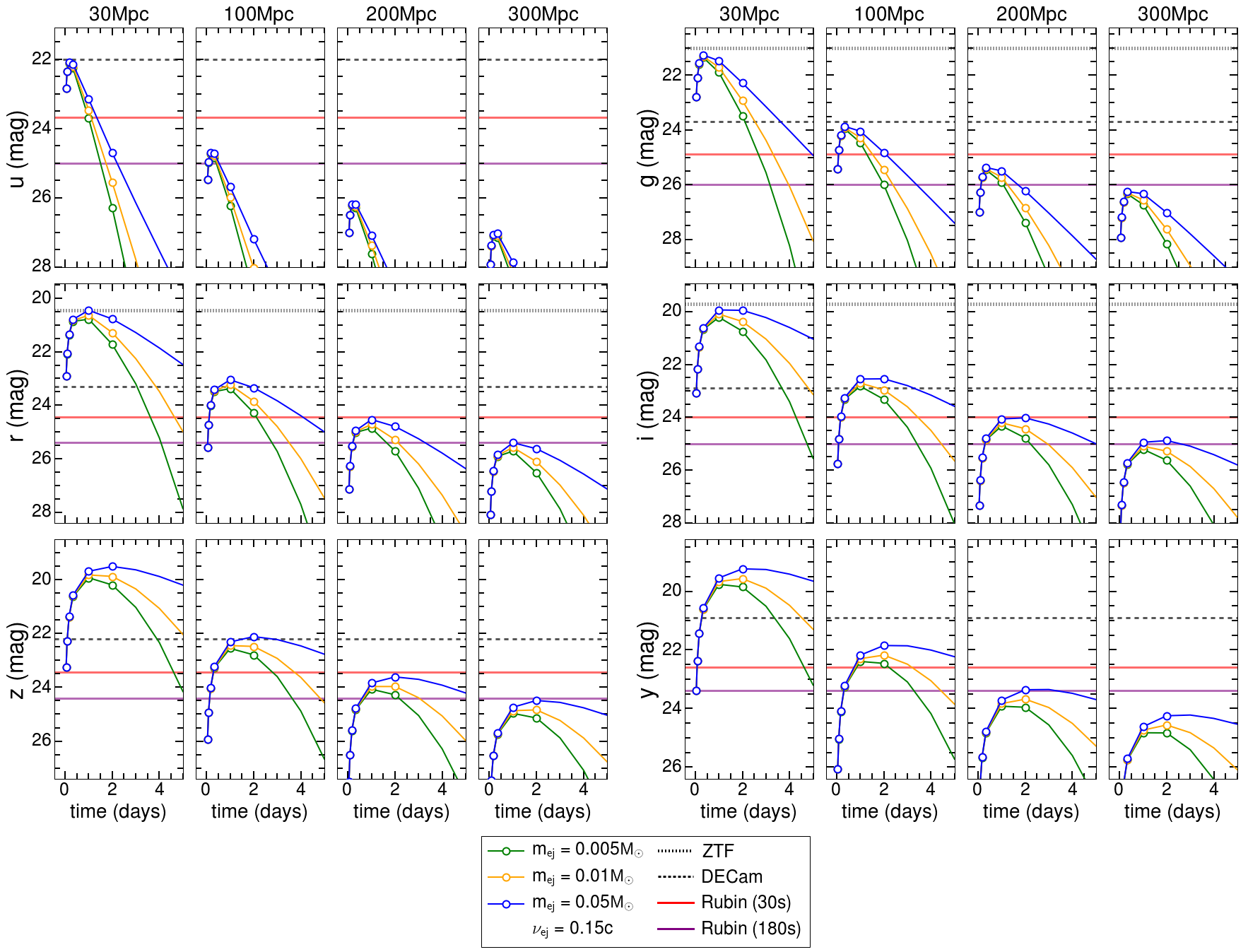}}
\caption{Same as Figure\,\ref{Fig:KNredblue}, but only the red KN component is considered to simulate the light curves, instead of a combination of a red and a blue component. The KN light-curves are adapted from models described in \cite{Villar17}. Since NS--BH mergers are expected to be accompanied by such redder KNe at larger distances, a longer and deeper monitoring is preferred (see \S\ref{SubSubSec:NSBH}).}
 \label{Fig:KNred}
\end{figure*}

A second key strength of our proposed strategy builds on the unique capability of Rubin to map the faint end of the KN brightness distribution. Systematic GW follow-up during O3 made it possible to add limits on the intrinsic KN luminosity function \citep{Kasliwal2020}. However, the intrinsically faint end $M > -15$\,mag, expected, for example, when the ejecta mass is lower than GW170817, is still poorly probed \citep[but see][]{Gompertz2018}. Observationally faint KN emission can also result from the most distant NS--NS mergers detected by the GW interferometers. During O5, NS--NS mergers are expected to be detected out to beyond $\sim300$ Mpc \citep{PeSi2021}. 
As shown in Figure\,\ref{Fig:KNredblue} and \ref{Fig:KNred}, Rubin is the only survey instrument able to discover red KNe at those distances. 

Deep, rapid multi-band observations are a crucial aspect of EM follow-up of NS--NS mergers as: (i) the blue KN component is not guaranteed to be present in all NS--NS mergers \citep{2015MNRAS.446.1115M}; (ii) even if present, the brightness of the blue KN component is angle-dependent, and will thus depend on our line of sight to the NS--NS merger \citep[e.g.,][]{Kasen17, Bulla2019, Nativi2021a}.  A solid discovery strategy of EM counterparts to NS--NS mergers has thus to be built around the capability to detect the red KN component.  As shown in Figure~\ref{Fig:KNred}, the red emission from KNe at $200$ Mpc and with small ejecta mass  $M_\mathrm{ej,red}=0.005\,\rm{M_{\odot}}$ \citep[$\sim$ one order of magnitude less than the ejecta mass inferred for the KN associated with GW170817, e.g.][]{Cowperthwaite17,Drout17,Kasliwal17, Pian2017, Smartt17,Villar17}
is well within the reach of one Rubin visit, while it is beyond or at the very limit of what other instrument surveys in the Southern hemisphere will be able to detect. Of those, DECam is the most sensitive, however its FoV is about 1/3 of Rubin's ant it lacks all-sky reference images for image subtraction. Rubin observations of KNe will allow us to probe the \emph{diversity} of the ejecta properties of NS--NS mergers in ways that are simply not accessible otherwise \citep[but see works that present the KN diversity based on short GRB observations, for example][]{Gompertz2018, Ascenzi2019, Lamb2019, Troja2019KN, Rossi2020}.

Set to start in late 2024 or early 2025, O5 will bring radical improvements in the detection of compact object coalescences. KAGRA and Virgo are expected to approach design sensitivity  (130\,Mpc and 150–-260\,Mpc, respectively) by 2025 and the orientation-averaged range of the LIGO detectors, with A+ upgrade, will be as large as 330\,Mpc for NS--NS mergers. Localizations can therefore become extremely well constrained with $\Omega_{\rm{90\%}}<$20 deg$^{2}$ out to $\sim 150$\,Mpc during Rubin operations. Given current NS--NS rates, we should expect 9--90\,events\,yr$^{-1}$ during O5 with localization regions smaller than 100\,deg$^2$ \citep[Table\,\ref{tab:detection_prospects};][]{PeSi2021}. These improvements are expected to greatly increase the number of well-localized mergers from O4 (see Table\,\ref{tab:detection_prospects}). Importantly, deep questions regarding GW sources cannot be solved during O4 with the small number of counterparts expected to be found with current facilities, thus they will remain open questions in the LSST era.

\begin{figure*}
\center{\includegraphics[scale=0.7]{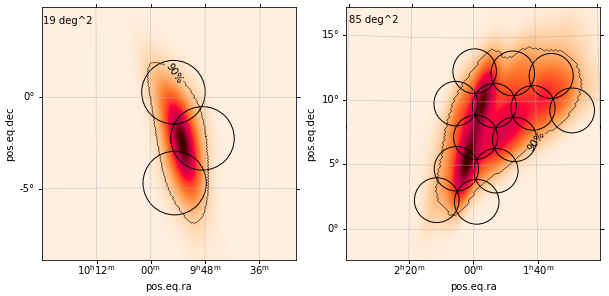}}
\caption{
Example of Rubin tiling of simulated GW skymaps for NS--NS mergers localized within $\Omega_{\rm{90\%}} < 20\,\rm{deg^2}$ (left panel) and $20\,\rm{deg^2}<\Omega_{\rm{90\%}}\le100\,\rm{deg^2}$ (right panel). The tiling pattern was created using \texttt{gwemopt} \citep{CoSt2016a} to include most ($\gtrsim 90\%$) of the integrated localization probability. We expect most skymaps with $\Omega_{\rm{90\%}} < 20\,\rm{deg^2}$ to require four Rubin pointings or less to cover $>90\%$ of the probability area, accounting for small offsets to be applied between exposures to cover chip gaps.}
 \label{fig:tiling}
\end{figure*}

In this work we design our strategies based on the expected performance of the GW detectors in O5 \citep{Abbott2020LRRprospects, PeSi2021}.
For some NS--NS mergers, Rubin can thus image the entire localization region with a relatively small number of pointings (Figure\,\ref{fig:tiling}), with dithered pointings that will be needed to cover chip gaps. This implies that Rubin will be able to capture the multi-band evolution of KNe potentially starting as early as minutes after the GW trigger. The earliest on-source time will be dictated by the position of the target in the sky for most events.

\begin{figure*}
\center{\includegraphics[width=\textwidth]{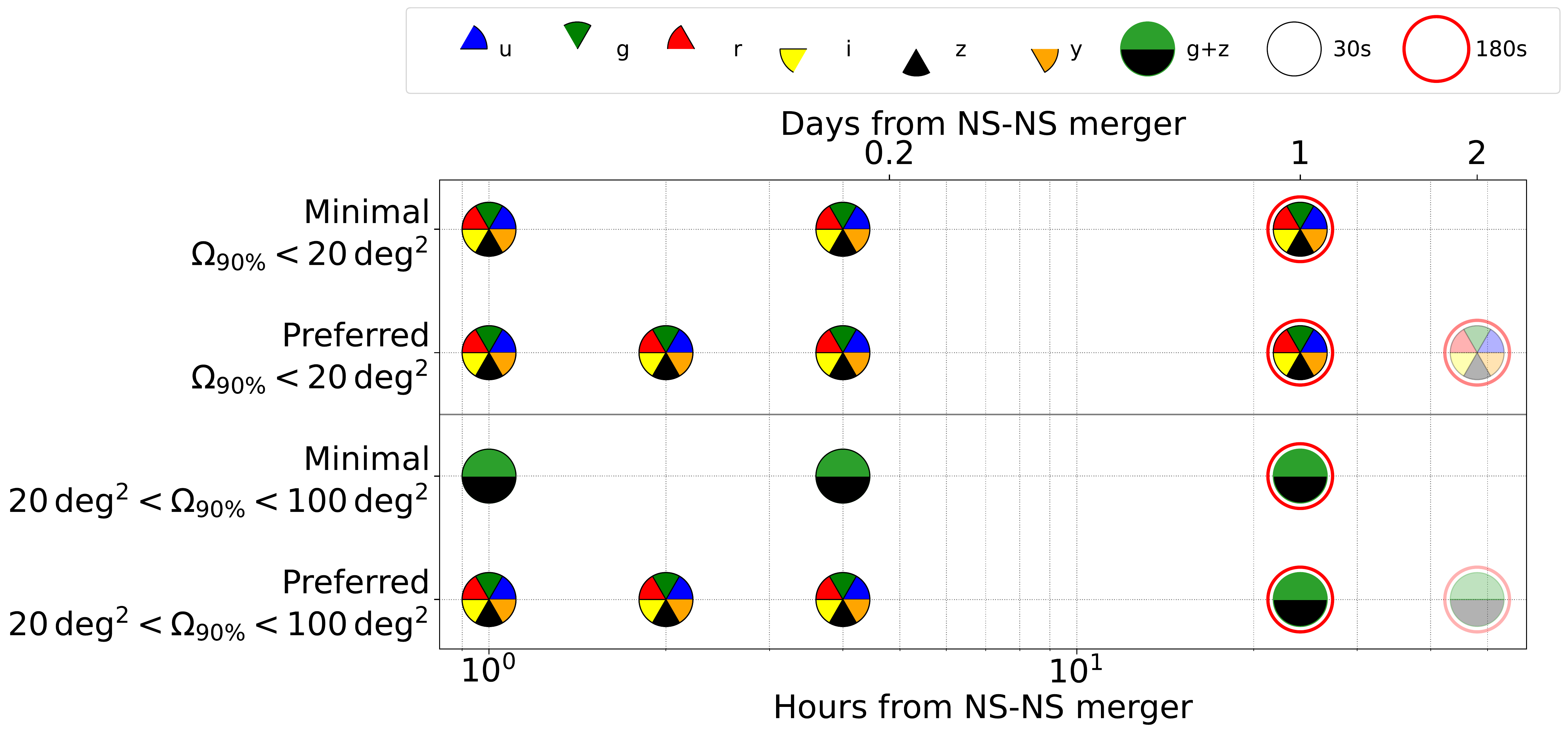}}
\center{\includegraphics[width=\textwidth]{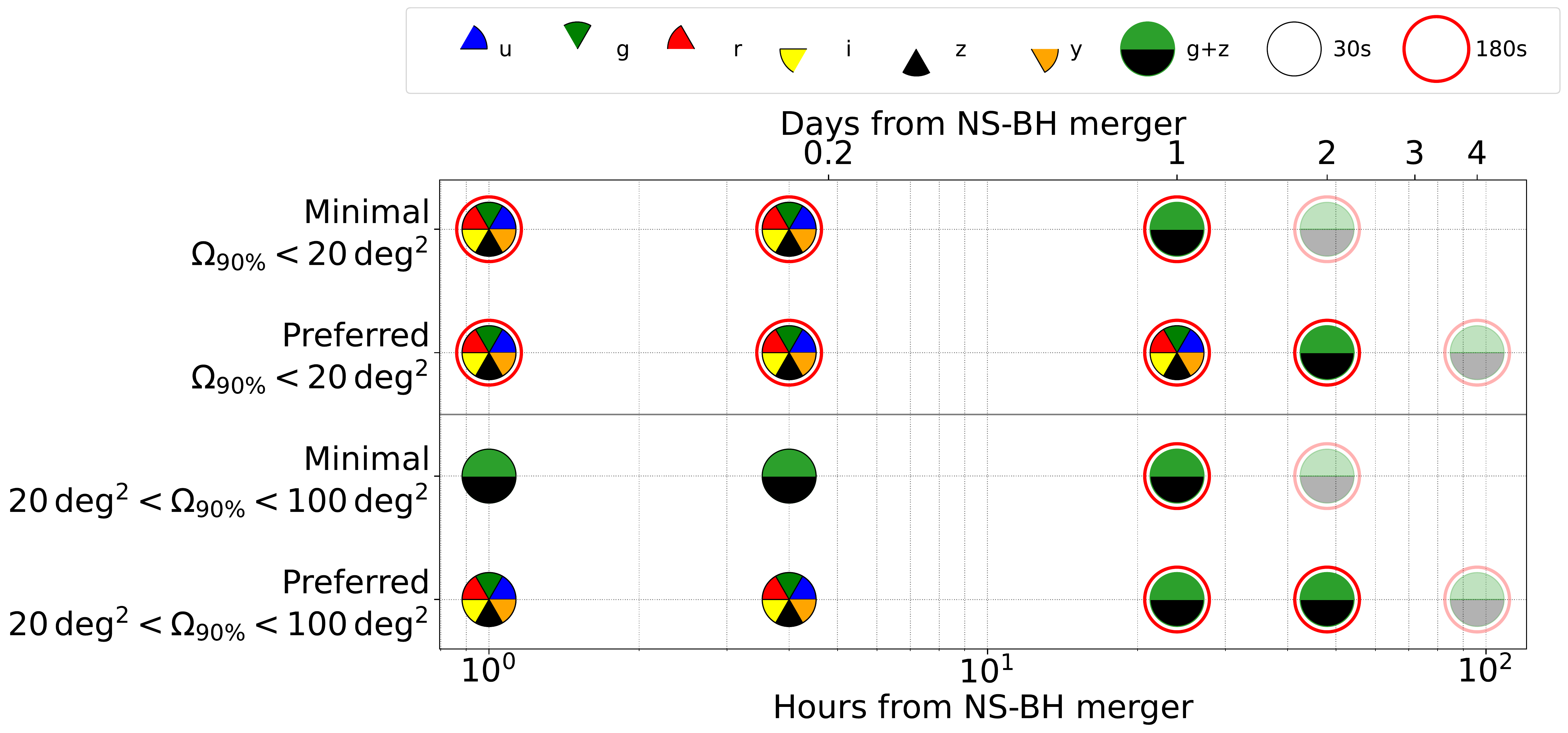}}
\caption{
Rubin observational cadences for NS--NS (top) and NS--BH (bottom) mergers follow-up with Rubin. Observations in $g+z$ filters will be replaced by observations in $g+i$ filters during bright dark time due to the limit of five filters available each night. For NS--NS mergers, we envision 30\,s exposures in each filter on the first night, 180\,s exposures (markers circled in red) on the following nights. For NS--BH mergers, which are expected to be found at larger distances, 180\,s exposures should be employed from the first night. Solid markers indicate planned observations over the entire localization area, while semi-transparent markers indicate possible extra observations to be carried out if the optical counterpart has not yet been identified.
}
 \label{Fig:gantt_ns}
\end{figure*}

Below, we outline our {\it minimal} and {\it preferred} Rubin ToO observing strategies of NS--NS mergers adopting an event rate of $R_{\text{BNS}} = 286^{+510}_{-237}$\,Gpc$^{-3}$yr$^{-1}$ for the median and 90\% symmetric credible intervals \citep{GWTC-2.1arXiv}. The time budget for ToO follow-up is calculated based on the expected GW event discovery rates for O5 \citep[Table\,\ref{tab:detection_prospects};][]{PeSi2021}. The observing strategies are summarized in Figure\,\ref{Fig:gantt_ns} and Table~\ref{tab:strategies}.

\textbf{\textit{Minimal} strategy:} On the first night, we propose at least two 5-filter visits ($u+g+r+i+y$ in dark time and $g+r+i+z+y$ in bright time; 30\,s exposure time for each filter) of well-localized NS--NS mergers with $\Omega_{\rm{90\%}}\le20\,\rm{deg^2}$ and whose sky position and timing are favorable for prompt follow up (i.e. within hours since GW trigger). Continued follow up during the first night is desirable, as outlined in the {\it preferred} strategy.  The bluer $u$ and $g$ bands are of particular interest as there are predictions of a free-neutron decay pulse within the first few hours after merger \citep{2015MNRAS.446.1115M}. We will aim at obtaining epochs at 1\,hr and 4\,hr, with a larger time spacing if observing conditions allow it. These observations will allow us to  identify new transients and separate KN candidates from background supernovae by measuring rapid luminosity and color evolution between the two epochs. 

Deeper 180\,s observations should be obtained on the following night, approximately 1 day from the merger. The $5\sigma$ magnitude limits for 30\,s and 180\,s exposures are shown in Figure\,\ref{Fig:KNredblue}-\ref{Fig:KNred} (a correction for image subtraction noise, which depends on the projected distance from the host among other factors, is not applied). In particular, for 180s exposures we will reach $m_g^{\rm{lim}}\sim26$ mag,  $m_z^{lim}\sim24.4$ mag (ideal observing conditions, dark sky), corresponding to absolute magnitudes $M_g^{\rm{lim}}= -10.5$ and $M_z^{\rm{lim}}= -12.1$ at 200\,Mpc. Deep 180\,s epochs will enable the measurement of the likely rapid KN decay, which is crucial to distinguish KNe from other Galactic and extragalactic transients.

Based on the results from \cite{Cowperthwaite18, Andreoni2019S190510g}, observations in $g+z$ bands can be particularly effective at finding optical counterparts to NS--NS mergers, especially after the possible blue component fades away within $\sim$hours from the merger. The $g+z$ filter combination can sample the widest possible range of the EM spectrum while maximizing the sensitivity of the observing campaign of less well localized targets, for example avoiding the throughput losses of the $u$ and $y$ filters. However, the $g+i$ was also demonstrated to be an effective combination  \citep[e.g.,][]{Andreoni2019PASP}. We therefore suggest to employ $g+z$ observations ($g+i$ in dark time, when the $z$ filter is unavailable) of more coarsely localized events with $20\,\rm{deg^2}<\Omega_{\rm{90\%}}\le100\,\rm{deg^2}$, with the same cadence and exposure times as above.

On average, we anticipate that $N=4$ ($N=20$) Rubin pointings will be needed to cover the localization area of mergers with $\Omega_{\rm{90\%}}\le20\,\rm{deg^2}$ ($\Omega_{\rm{90\%}}\le100\,\rm{deg^2}$), see Figure\,\ref{fig:tiling}. With this strategy, we expect to spend $\sim1.85$\,hr ($\sim3.00$\,hr) per NS--NS merger with $\Omega_{\rm{90\%}}\le20\,\rm{deg^2}$ ($20\,\rm{deg^2}<\Omega_{\rm{90\%}}\le100\,\rm{deg^2}$).  

Based on the results obtained by \cite{PeSi2021} and reported in Table\,\ref{tab:detection_prospects}, the number of mergers with $\Omega_{\rm{90\%}}\le20\,\rm{deg^2}$ is $N = 13^{+29}_{-9.1}$\,yr$^{-1}$ during O5
($N = 22^{+49}_{-15}$\,yr$^{-1}$ for for $20\,\rm{deg^2}<\Omega_{\rm{90\%}}\le100\,\rm{deg^2}$) assuming a duration of 1 year.
These values translate to $\sim 7^{+15}_{-5}$ ($\sim 12^{+8}_{-3}$) sources accessible to Rubin.

Accounting for 7 well-localized mergers with $\Omega_{\rm{90\%}}\le20\,\rm{deg^2}$ and 3 particularly promising mergers with $20\,\rm{deg^2}<\Omega_{\rm{90\%}}\le100\,\rm{deg^2}$ (chosen depending on GW signal-to-noise ratio, distance, observability, distance from the moon and the Galactic plane) to be followed up, the desired time allocation for NS--NS mergers is about 13\,hr and 9\,hr, respectively, during O5, for a total of $\sim 22$\,hr.

\begin{table*}[ht!]
    \centering
    \begin{tabular}{@{\extracolsep{4pt}}ccccccc}
    \hline\hline
    & \multicolumn{3}{c}{O4} & \multicolumn{3}{c}{O5} \\
\cline{2-4} \cline{5-7}
         & Total & $20<\Omega_{\rm{90\%}}\le100$ & $\Omega_{\rm{90\%}}\le20$ & Total & $20<\Omega_{\rm{90\%}}\le100$ & $\Omega_{\rm{90\%}}\le20$  \\ 
    \hline
  NS--NS & $34^{+78}_{-25}$ & $2.5^{+5.7}_{-1.8}$ & $2.4^{+5.6}_{-1.8}$ & $190^{+410}_{-130}$ & $22^{+49}_{-15}$ & $13^{+29}_{-9.1}$ \\
  NS--BH & $72^{+75}_{-38}$ & $6.8^{+7.1}_{-4.0}$ & $4.3^{+4.5}_{-2.5}$ & $360^{+360}_{-180}$ & $45^{+45}_{-23}$ & $23^{+23}_{-12}$ \\
  BH--BH & $106^{+65}_{-42}$ & $19^{+12}_{-7.7}$ & $15^{+9.3}_{-6.0}$ & $480^{+280}_{-180}$ & $104^{+61}_{-39}$ & $70^{+41}_{-26}$ \\
    \hline
    \end{tabular}
    \caption{Realistic expectations for NS--NS, NS--BH and BH--BH merger detection during LVK O5, assuming a duration of one calendar year for the run. Information is also reported for O4 for comparison, to stress the big difference in the number of well localized sources expected between O4 and O5. The table indicates the total number of expected detections and those with localization uncertainty $\Omega_{\rm{90\%}}\le20\,\rm{deg^2}$ and  $20\,\rm{deg^2}<\Omega_{\rm{90\%}}\le100\,\rm{deg^2}$. The reported values are based on results obtained by \cite{PeSi2021}.}
    \label{tab:detection_prospects}
\end{table*}

\textbf{\textit{Preferred} strategy: } Three sets of
five filter observations ($u+g+r+i+y$ in dark time and $g+r+i+z+y$ in bright time; 30\,s for each filter) should be employed.
Observations will be log-spaced in time with focus on the first night the object is available to sample the very early KN evolution (see \S\ref{sec:intro} and \S\ref{SubSubSec:NSNS} regarding the scientific significance of rich observations within few hours from the merger) at 1\,hr, 2\,hr, and 4\,hr from all NS--NS mergers with
$\Omega_{\rm{90\%}}\le100\,\rm{deg^2}$ and for which the sky position and time are favorable for rapid follow-up with Rubin. Additional observations at 8\,hr are desired, too, if they are possible to perform.

On the second night, the entire localization area should be imaged with 180\,s exposures in
all five filters for events with $\Omega_{\rm{90\%}}\le20\,\rm{deg^2}$ and $g+z$ filters for events with  $20\,\rm{deg^2}<\Omega_{\rm{90\%}}\le100\,\rm{deg^2}$. 

If an optical counterpart has not been unambiguously identified, we suggest performing a final set of observations on the third night. This could be the only way of effectively distinguishing a KN from supernovae and other contaminant sources. 

With this {\it preferred} strategy, the average Rubin investment of time per NS--NS merger is 2.19\,hr (5.59\,hr) for GW sources localized within $\Omega_{\rm{90\%}}\le20\,\rm{deg^2}$ ($20\,\rm{deg^2}<\Omega_{\rm{90\%}}\le100\,\rm{deg^2}$).

Accounting again for 7 well-localized mergers with $\Omega_{\rm{90\%}}\le20\,\rm{deg^2}$ and the best 3 mergers with $20\,\rm{deg^2}<\Omega_{\rm{90\%}}\le100\,\rm{deg^2}$ to be followed up, the desired time allocation for NS--NS mergers is 15.32\,hr and 16.78\,hr, respectively, during O5, for a total of about 32\,hr. 

We stress that the 10\,hr budgeted for the {\it preferred} strategy more than in the {\it minimal} strategy can add great scientific value by providing multi-band, highly-cadenced data that will make KN discovery more robust, but will also allow us to measure with precision the temperature evolution of the short-lived, elusive blue component.  This will be precious especially if the number of detected NS--NS mergers in GWs is similar to, or lower than, the median expected value (Table\,\ref{tab:detection_prospects}). The {\it preferred} strategy will also be more effective at separating KNe from un-related transients photometrically in real time. Future work is planned to evaluate the impact of those strategies on parameter estimation for a set of KN models and further optimize them \citep[see for example][]{Sravan2021arXiv}. Future analysis could also evaluate the implementation of hybrid strategies in which, for the same trigger, higher probability regions and low probability regions are tiled with a different cadence or filter choice.

%---------------------------------------------------------
\subsubsection{The Rubin quest for the unknown: EM counterparts to NS--BH mergers:}
\label{SubSubSec:NSBH}

As of the end of O3, several GW detections of NS--BH merger candidates have been reported \citep{LIGO_GWTC2_pop_2020, Abbott2021NSBH}, but no EM counterpart to a NS--BH  merger was found \citep[e.g.,][for the most robust NS--BH GW detections]{AnCo2020, Dichiara2021arXiv}. Extensive follow-up was also performed for the peculiar source GW190814 \citep{Dobie2019, Gomez2019, Ackley2020, Andreoni2020S190814bv, Morgan2020, Thakur2020, Vieira2020, Watson2020, Alexander2021arXiv, deWet2021, Dobie2021arXiv, Kilpatrick2021arXiv, Tucker2021arXiv}. However, the nature of the secondary component of GW190814 is unclear, as it can be either the lightest black hole or the heaviest neutron star ever discovered in a double compact-object system \citep{Abbott2020GW190814}.

Yet, some NS--BH mergers are expected to be accompanied by KN emission not dissimilar in nature from the KN emission from NS--NS mergers. Their GW localizations are also expected to be similar to those of NS--NS mergers, despite their larger distance due to the larger amplitudes of their GW signals. 
The range of dynamical ejecta mass produced by NS--BH mergers is broad: it can be much less than in NS--NS mergers if the system lacks a fast spinning black hole or a very favourable mass ratio, but it might be up to $\sim10$ times larger than in NS--NS mergers, which would lead to luminous KNe peaking $\sim 1$ magnitude brighter than GW170817 \citep[e.g.,][]{KaFe2015,2015MNRAS.446.1115M, Bulla2019, Barbieri2020, Hotokezaka2020}. However, the amount of lanthanide-poor ejecta is expected to be low and, differently from NS--NS mergers, no neutron precursor is expected at early times \citep{2015MNRAS.446.1115M}. While some early blue emission from the disk winds is not excluded, the general expectation is that KNe associated to NS--BH mergers will be typically dominated by the near-IR component. 

Especially in the case of NS--BH mergers, the deep sensitivity of Rubin brings an additional advantage compared to all the other survey instruments. GW detectors are sensitive to NS--BH mergers at distances extending to several hundred Mpc, which implies that, on average, NS--BH mergers will be localized at larger distances than NS--NS mergers \citep{Abbott2020LRRprospects} (factor of a few). 
The larger distances of NS--BH systems detected through their GW emission cancel out the advantage of their intrinsically more luminous KN emission. NS--BH mergers will be thus on average observed as fainter signals in the EM spectrum and will greatly benefit from the Rubin large collecting area.

The strategies chosen for NS--BH mergers envision at least two sets of observations on the first night from the merger, followed by follow-up until a few days later. Deep observations since the beginning of follow-up campaigns will probe the emission at early times even for distant events. A longer monitoring time is likely going to be required in order to recognise NS--BH KNe, which might evolve slower than GW170817, hence being harder to distinguish from supernovae and other types of unrelated transients.

In addition to the unknown light curve behavior, a major source of uncertainty is the intrinsic rate of NS--BH mergers in the local universe, which is constrained by GW observations as $R_{NS-BH}={45}_{-33}^{+75}\,{\mathrm{Gpc}}^{-3}\,{\mathrm{yr}}^{-1}$  \citep[assuming that GW200105 and GW200115 are representative of the NSBH population, or  $R_{NS-BH}={130}_{-69}^{+112}\,{\mathrm{Gpc}}^{-3}\,{\mathrm{yr}}^{-1}$ assuming a broader distribution of component masses;][]{Abbott2021NSBH}, still consistent with the 90\%  confidence range of NS--NS merger rate $R$=80--810\,$\rm{Gpc^{-3}yr^{-1}}$ \citep{LIGO_GWTC2_pop_2020}. The observing strategies for NS--BH follow-up are also summarized in Figure\,\ref{Fig:gantt_ns} and Table~\ref{tab:strategies}.

\textbf{\textit{Minimal} strategy:} For well localized events with $\Omega_{\rm{90\%}}\le20\,\rm{deg^2}$, two sets of deep five-filter observations ($u+g+r+i+y$ in dark time and $g+r+i+z+y$ in bright time; 180\,s exposure time) should be carried out at 1\,hr and 4\,hr from the merger on the first night. On the second night, $g+z$ ($g+i$ in dark time) exposures (again 180\,s) should follow if a counterpart is not yet identified. An additional observation pair on the third night might be desired if the counterpart remains elusive.

More coarsely localized mergers with $20\,\rm{deg^2}<\Omega_{\rm{90\%}}\le100\,\rm{deg^2}$ will be observed with the same cadence, but observations should be carried out only in $g+z$ ($g+i$) filters. Only the closest of such coarsely localized events should be followed up (for instance, within a luminosity distance of 250\,Mpc, where a faint KN peaking at $\sim -13$ would be observable at $\sim 24$\,mag). Hence we suggest observations to be carried out with 30\,s exposure times on the first night, then 180\,s on the following nights to detect possible rapidly fading transients.  

The average LSST investment of time per NS--BH merger in the {\it minimal} strategy is $\sim2.79$\,hr and $\sim 3.00$\,hr per NS--BH merger with $\Omega_{\rm{90\%}}\le20\,\rm{deg^2}$ and $20\,\rm{deg^2}<\Omega_{\rm{90\%}}\le100\,\rm{deg^2}$, respectively.

Based on the results presented in \cite{PeSi2021} and summarized in Table\,\ref{tab:detection_prospects}, we can expect $\sim 12^{+12}_{-6}$ ($\sim 24^{+24}_{-12}$) NS--BH mergers to be accessible for Rubin that are localized within $\Omega_{\rm{90\%}}\le20\,\rm{deg^2}$ ($20\,\rm{deg^2}<\Omega_{\rm{90\%}}\le100\,\rm{deg^2}$) in O5. Considering 12 well-localized events and 3 particularly promising events that are more coarsely localized, the total time allocation for the {\it minimal} strategy to follow up NS--BH mergers with Rubin would be 33.48\,hr and 9.00\,hr for the two localization categories, respectively, for a total of 42.48\,hr.

\textbf{\textit{Preferred} strategy:}  
At least two sets of deep five-filter observations ($u+g+r+i+y$ in dark time and $g+r+i+z+y$ in bright time; 180\,s exposure time) should be obtained on the first night at 1\,hr and 4\,hr from the merger for all sources localized better than $\Omega_{\rm{90\%}}\le20\,\rm{deg^2}$. Additional data taken at 2\,hr and/or 8\,hr from the merger could help characterize the very early emission as outlined in \S\ref{SubSubSec:NSNS} for NS--NS mergers. 

For particularly well-localized NS--BH mergers ($\Omega_{\rm{90\%}}<20\,\rm{deg^2}$), the entire area should be imaged again on the second ($u+g+r+i+y$ or $g+r+i+z+y$ filters) and third ($g+z$ or $g+i$ filters) night, with exposures of 180\,s per filter. An additional epoch on the fourth night is desirable if a counterpart is yet to be unambiguously identified. This systematic approach may be necessary to obtain a uniform dataset to recognize and characterize possible yet unknown counterparts to NS--BH mergers in an unbiased way.

More coarsely localized mergers with $20\,\rm{deg^2}<\Omega_{\rm{90\%}}\le100\,\rm{deg^2}$ will be observed with the same cadence, but observations should be carried out only in $g+z$ filters from the second night onward. Since only the closest (e.g. $D < 250$\,Mpc, see above) of such coarsely localized events should be followed up, the exposure time should be of 30\,s on the first night and 180\,s on the following nights. 

The average LSST investment of time per NS--BH merger in the {\it preferred} strategy is 3.97\,hr and 4.43\,hr per NS--BH merger with $\Omega_{\rm{90\%}}\le20\,\rm{deg^2}$ and $20\,\rm{deg^2}<\Omega_{\rm{90\%}}\le100\,\rm{deg^2}$, respectively.

Considering again 12 well-localized NS--BH events and 3 particularly significant and nearby events that are more coarsely localized, the total time allocation for the {\it preferred} strategy to follow up NS--BH mergers with Rubin would be 47.64\,hr (13.29\,hr) for events localized within $\Omega_{\rm{90\%}}<20\,\rm{deg^2}$ ($20\,\rm{deg^2}<\Omega_{\rm{90\%}}\le100\,\rm{deg^2}$) in O5, for a total of about 61\,hr.

%---------------------------------------------------------
\subsubsection{The Rubin quest for the unknown: EM counterparts to BH--BH mergers}
Theoretical speculations on EM counterparts to BH--BH mergers experienced a surge of interest because of the possible association of a burst of $\gamma$-rays detected by the Fermi satellite with the BH--BH merger event GW150914 \citep{Connaughton16} and the discovery of an AGN flare that might be associated with GW190521 \citep{Graham2020PhRv}.

Follow-up observations of BH--BH mergers are also extremely valuable to probe formation channels of LVK stellar black holes, even in the case of non-detection or multiple potential associations. In the case of BH--BH mergers inducing AGN flares, following up the better localized events as described here can produce a constraint on the fraction of BH--BH mergers happening in AGN disks with 2-3 orders of magnitude less events than without a follow-up, and simultaneously produce cosmological results more constraining than standard sirens without a counterpart \citep{Palmese2021}.

BH--BH mergers are routinely detected by the detectors through their GW emission, but to date an unambiguous association with an EM counterpart is still missing. Theoretical models of EM counterparts from BH--BH mergers are highly speculative and span a wide range of possible morphologies \citep{Perna,Loeb,Stone,deMink,McKernan}. On the observational side, few deep follow-up campaigns were performed to date \citep[e.g.,][for S191216ap]{Bhakta2021}, the most complete being dedicated to observations of the well-localized event GW190814, if the progenitor system was indeed a BH--BH binary (see \S\ref{SubSubSec:NSBH}).  Since no viable counterpart was found, the existence and properties of EM transient emission from BH--BH mergers is still a completely open question in astrophysics.  Given the current large uncertainty of possible EM counterparts, we design a model-agnostic Rubin observational strategy of two nearby, very well localized BH--BH mergers. 

\begin{figure*}[t!]
\center{\includegraphics[width=0.8\textwidth]{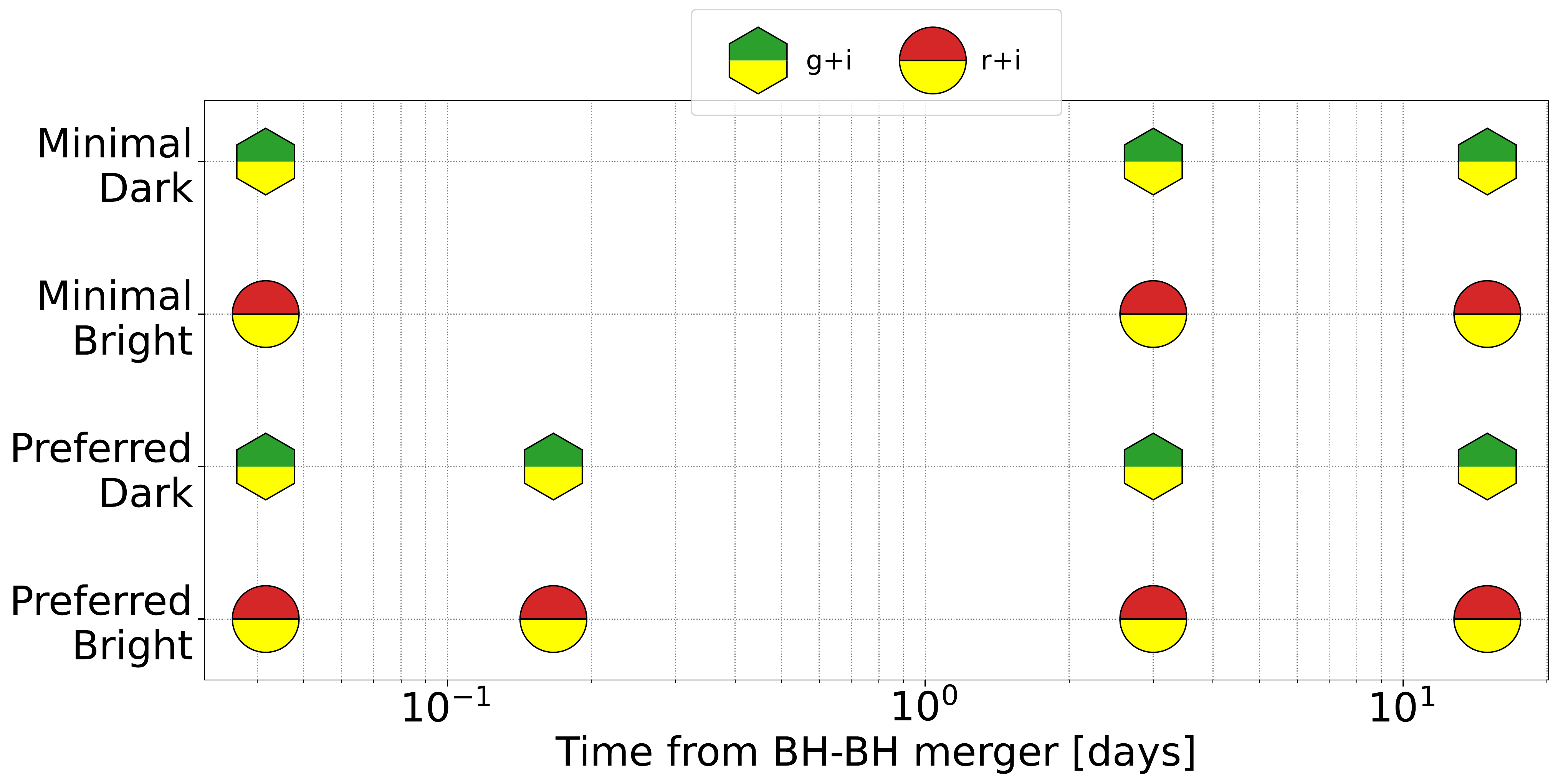}}
\caption{
Cadences for BH--BH merger follow-up with Rubin. In dark time, observations in $g$+$i$ filters are preferred to $r$+$i$ observations, to be carried out in bright time.
}
 \label{Fig:gantt_bhbh}
\end{figure*}

 The observing strategies are summarized in Table~\ref{tab:strategies} and in Figure\,\ref{Fig:gantt_bhbh}. We note that, thanks to the large number of expected BH--BH mergers in O5, localization regions of many BH--BH mergers could be probed by the LSST WFD survey. However, equipped with ToO capabilities, Rubin will probe the existence and properties of transients from BH--BH mergers at short and medium time scales with unparalleled sensitivity among ground-based surveys, thus opening up a completely new window of investigation on our Universe.

\textbf{\textit{Minimal} strategy:} For Rubin follow up of promptly accessible (i.e. within hours of GW detection) BH--BH mergers at $d_L\le500$ Mpc  with $\Omega_{90\%}\le15\,\rm{deg^2}$. The expected rate of well-localized BH--BH mergers is high (Table\,\ref{tab:detection_prospects}), hence several could be also detected at low distances. Given the loudness of these events, we expect to be able to cover the GW localization region of well-localized BH--BH mergers with only two or three Rubin pointings. Given that properties of transient counterparts to BH--BH mergers are unknown, we advocate for follow up in filter pairs that maximize depth for our search for EM counterparts and, possibly, sample well separated regions of the optical spectrum.  We propose deep  $g+i$ observations during dark time and $r+i$ observations during bright time ($180$\,s exposure for each filter), which will bring the highest throughput. 

We propose deep $g+i$ (or $r+i$ during bright time) observations (180\,s exposure for each filter) at 1 hr, 3 days and 15 days after the merger. The average investment of Rubin time per BH--BH merger is $ 0.72$\,hr (total of 1.45 hr $\rm{yr^{-1}}$). For a 180\,s exposure observation anticipate reaching a $5\sigma$ magnitude limit  $m_g^{\rm{lim}}\sim26$\,mag  $m_i^{\rm{lim}}\sim25$\,mag (under ideal conditions of dark sky and zenith pointing), corresponding to absolute magnitudes $M_g^{\rm{lim}}= -12.5$\,mag and $M_i^{\rm{lim}}= -13.5$\,mag at 500\,Mpc.

\textbf{\textit{Preferred} strategy:} Same as the the {\it minimal} strategy outlined above, but with the addition of another epoch of deep $g+i$  observations  (or $r+i$ during bright time) during the first night. This strategy will allow us to map the very short time-scales of variability of potential EM transients associated with BH--BH mergers, as well as the longer time scales of evolution of $\sim$\,weeks.  The average investment of Rubin time per BH--BH merger is $0.96$\,hr (total of about 1.83\,hr $\rm{yr^{-1}}$).

With the BH--BH follow-up campaign described here, under ideal observing conditions, Rubin will extend the discovery space $\sim3$ magnitudes deeper than previous campaigns, probing fast and slow time scales of evolution of EM counterparts to BH--BH mergers in two bands (hence providing color information). The key advantage of the {\it preferred} strategy, compared to the {\it minimal} strategy, is the capability to sample the very short timescales of evolution of the transients.

%---------------------------------------------------------
\subsubsection{The Rubin quest for the unknown: unmodeled GW sources}
\label{SubSubSec:GWunknown}
This class of GW triggers include sources found through GW unmodeled source pipelines, which are not necessarily of compact-object mergers in origin and might include very nearby supernova explosions and things we may not even have thought of. 

Only one (poorly localized) candidate of such events was found to date, thus we consider Rubin follow-up of one unidentified GW source during O5, with localization $\Omega_{\rm{90\%}}\le 100$\,deg$^2$. We expect to be able to cover the localization region with $\sim20$ Rubin pointings.  We propose $g+z$ 30s-exposure observations during the first night, at 3 days and 15 days to sample the EM spectrum with deep sensitivity ($r+i$ will be used during dark time). For GW sources for which the entire region can be covered at low airmass, two $g+z$ (or $g+i$ in dark time) epochs will be acquired during the first night. With this strategy, we will be able to constrain the presence of EM counterparts to unidentified GW sources across the spectrum, both on short (i.e. intra-night) and longer time-scales of weeks.  The average investment of time per GW trigger is $1.33$\,hr. This is a small investment of Rubin time, which holds promises for high discovery potential and significant scientific impact. The observing strategy is summarized in Table~\ref{tab:strategies}.

\begin{sidewaystable*}
\vspace{4in}
    \centering
    \begin{tabular}{r|l|l|l|}
        \toprule
         & \textbf{{\it minimal} Strategy NS--NS} & \textbf{{\it preferred} Strategy NS--NS} \hspace{.3in}\\
        \midrule
        \textbf{Sequence} & $(u)grizy$ (30s) at 1h, 4h and $gz(i)$ (180s) at 24h for $\Omega_{\rm{90\%}}\le20$ deg$^2$ & $(u)grizy$ (30s) at 1h, 2h, 4h and  $gz(i)$ (180s) at 24h\\ 
        & $gz(i)$ (30s) at 1h, 4h and $gz(i)$ (180s) at 24h for $20$ deg$^{2}<\Omega_{\rm{90\%}}\le100$ deg$^2$ & $(u)grizy$ (30s) at 1h, 2h, 4h and  $gz(i)$ (180s) at 24h\\ 
        \textbf{Tiles} & 4 for $\Omega_{\rm{90\%}}\le20$ deg$^2$ & 4 \\
         & 20 for $20$ deg$^{2}<\Omega_{\rm{90\%}}\le100$ deg$^2$& 20\\
        \textbf{Average Time per Event} & 1.85 hr for $\Omega_{\rm{90\%}}\le20$ deg$^2$ &  2.19 hr\\
        & 3.0 hr for $20$  deg$^{2}<\Omega_{\rm{90\%}}\le100$ deg$^2$& 5.59 hr \\
        \textbf{Events per year} & Same as {\it preferred} &  7 with $\Omega_{\rm{90\%}}\le20$ deg$^2$\\
        &Same as {\it preferred} & 3 $20$ deg$^{2}<\Omega_{\rm{90\%}}\le100$ deg$^2$ \\
        \textbf{Total Time per LSST year} & 21.95\,hr & 32.1\,hr\\
          \toprule
       & \textbf{{\it minimal} Strategy NS--BH} & \textbf{{\it preferred} Strategy NS--BH} \hspace{.3in}\\
               \toprule
\textbf{Sequence} & \textit{(u)grizy} (180s) at 1h, 4h and \textit{gz(i)} at 24h (180s) for $\Omega_{\rm{90\%}}\le20$ deg$^2$& \textit{(u)grizy} (180s) at 1h, 4h, 24h and \textit{gz(i)} (180s) at 48h\\
& \textit{gz(i)} (180s) at 1h, 4h, 24h (180s) for $20$ deg$^{2}<\Omega_{\rm{90\%}}\le100$ deg$^2$& \textit{(u)grizy} (180s) at 1h, 4h and \textit{gz(i)} at 24h, 48h (180s)\\
        \textbf{Tiles} & 4 for $\Omega_{\rm{90\%}}\le20$ deg$^2$ & 4 \\
         & 20 for $20$ deg$^{2}<\Omega_{\rm{90\%}}\le100$ deg$^2$& 20\\
        \textbf{Average Time per Event} & 2.79\,hr for $\Omega_{\rm{90\%}}\le20$ deg$^2$ & 3.97\,hr for  \\
        & 3.0\,hr for $20$ deg$^{2}<\Omega_{\rm{90\%}}\le100$ deg$^2$ & 4.43\,hr for  \\
 \textbf{Events per year}& Same as {\it preferred} &12 with $\Omega_{\rm{90\%}}\le20$ deg$^2$ \\
        & Same as {\it preferred}&3 with $20$ deg$^{2}<\Omega_{\rm{90\%}}\le100$ deg$^2$ \\
\textbf{Total Time per LSST year}&  42.48\,hr & 60.93\,hr\\
          \toprule
      & \textbf{{\it minimal} Strategy BH--BH} & \textbf{{\it preferred} Strategy BH--BH} \hspace{.3in}\\
               \toprule
\textbf{Sequence} & $gi^{*}$ (180s ) at 1h, 3d, 15d & $gi^{*}$ (180s ) at 1h, 4h, 3d, 15d\\
        \textbf{Tiles} & 2 & 2 \\
\textbf{Average Time per Event} & 0.725\,hr & 0.96\,hr\\
\textbf{Events per year}& $\sim2$ & $\sim2$\\
\textbf{Total Time per LSST year}& 1.45\,hr & 1.93\,hr\\
         \toprule
      & \textbf{{\it minimal} Strategy Unidentified GW} & \textbf{{\it preferred} Strategy Unidentified GW} \hspace{.3in}\\
               \toprule
                    \textbf{Sequence} &Same as {\it preferred}& $gi^{*}$ (30s) at 1h, 3d, 15d\\
                    \textbf{Tiles} & 20 & 20 \\
     \textbf{Average Time per Event}&Same as {\it preferred}&1.33 hr\\
     \textbf{Events per year} &Same as {\it preferred}& 1\\
  \textbf{Total Time per LSST year}&  Same as {\it preferred} & 1.33 hr\\   
          \toprule
        \toprule
     & \textbf{{\it minimal} Strategy - Total Allocation} & \textbf{{\it preferred} Strategy  - Total Allocation} \hspace{.3in}\\
      & 67.2 hr ($\sim$1.39\% LSST time) & 96.2 hr ($\sim$1.99\% LSST time)\\
        \bottomrule
    \end{tabular}
    \caption{{\bf Summary of strategies and expected time allocations.} ($^*$) we will use $r$-band instead of $g$-band during bright time. Parentheses for filters indicate if a filter will be included instead of another depending on whether the observation is taken during dark or bright time, as outlined in \S \ref{subsec: high-level-description}}
        \label{tab:strategies}
\end{sidewaystable*}

%%%%%%%%%%%%%%%%%%%%%%%%%%%%%%%%%%%%%%%%%%%%%%%%%%%%%%%%%%%%%%%%%%%%%%%%%%%%%%%%%%%%%%%%%%
\section{Performance Evaluation}
\label{sec: performance}

As explained in \S\ref{subsec: high-level-description}, Rubin ToO observations are key to EM counterpart discovery in the next decade. If the {\it preferred} strategy outlined above is implemented, we expect an EM counterpart discovery in the vast majority of NS--NS mergers within a distance of $\lesssim 300$\,Mpc, assuming that GW170817 is not too dissimilar from the typical KN from NS--NS mergers. With the Rubin {\it minimal} ToO strategy, we anticipate a lower level of success (e.g., less timely EM candidate identification, which might prevent subsequent characterization of the source with smaller FoV facilities, or limited  information on the early time properties of the EM counterpart, which will preclude the identification of additional components of emission). Based on these considerations, we define a heuristic quantifier of the success of the ToO implementation for NS--NS merger follow-up as:
\begin{equation}
S_\mathrm{NS-NS} = \frac{(1+n_\mathrm{ep}+n_\mathrm{flt}+2 f_\mathrm{early})N_\mathrm{det}}{12 N_\mathrm{NS-NS}}
\label{eq:S_NSNS}
\end{equation}
where $N_\mathrm{NS-NS}$ is the number of NS--NS mergers detected by GW interferometers that satisfy the ToO activation criteria, $N_\mathrm{det}$ is the number of associated KN detections in Rubin ToOs, $n_\mathrm{ep}$ is average number of epochs per event in the strategy, $n_\mathrm{flt}$ is the average number of filters employed per event, and $f_\mathrm{early}$ is the fraction of the ToOs that lead to an identification of the counterpart within 1 day. This definition gives added value to higher cadence, multi-filter monitoring -- which is a requirement for an appropriate determination of the temperature evolution -- and to an early detection. The normalization of Eq.~\ref{eq:S_NSNS} is defined in such a way that a strategy that envisages observations in 5 filters, four epochs per event, and leads to the detection of all events within 1 day, yields $S_\mathrm{NS-NS}=1$. We note that many alternative, equally reasonable choices could have been made in defining such a metric: for example, a different weight could be assigned to $n_\mathrm{ep}$ and $n_\mathrm{flt}$ to emphasize higher cadence (or longer-lasting monitoring) with respect to an accurate determination of a smaller number of SEDs, or \textit{viceversa}. Still, such a change would not impact significantly our evaluation, neither quantitatively nor qualitatively.

\begin{figure*}[t]
    \centering
    \includegraphics[width=0.48\textwidth]{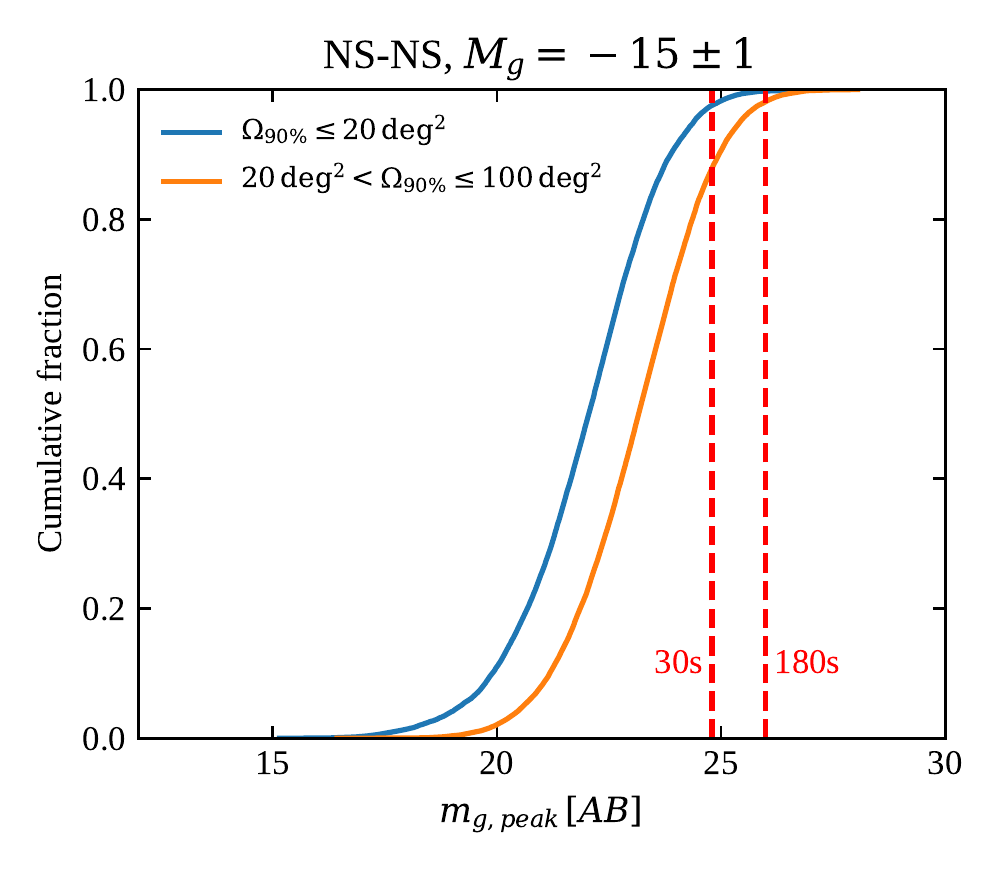}\includegraphics[width=0.48\textwidth]{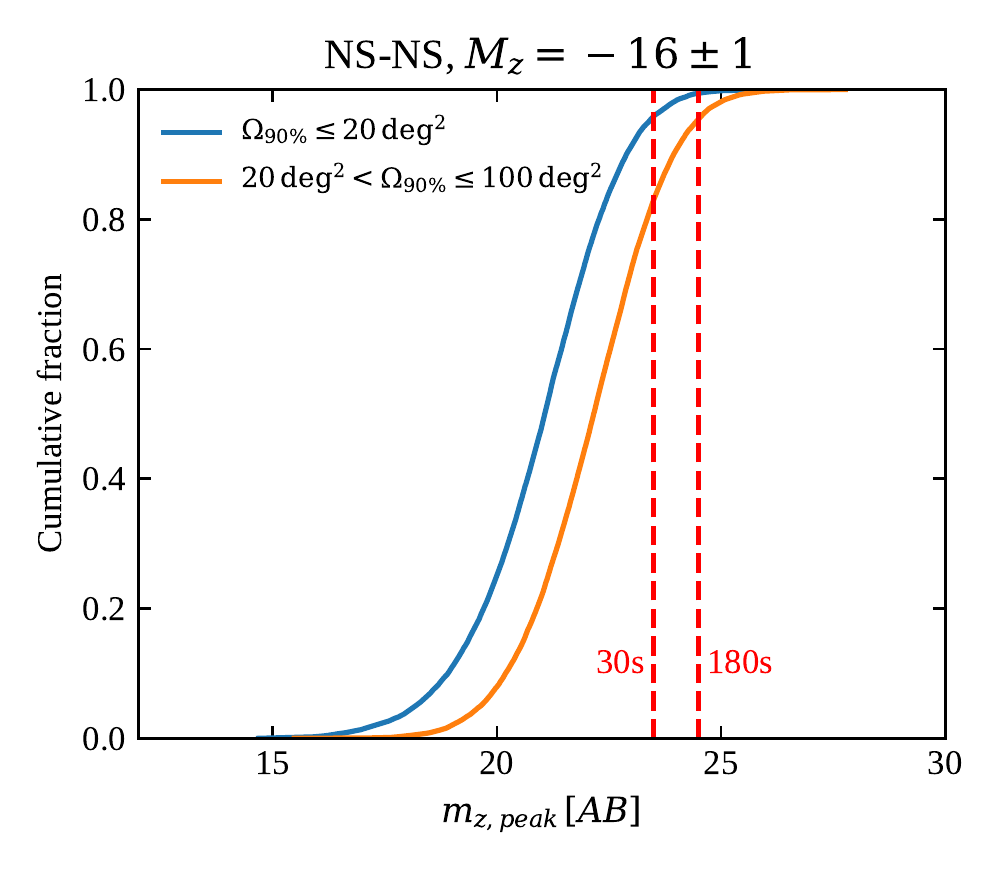}\\
    \includegraphics[width=0.48\textwidth]{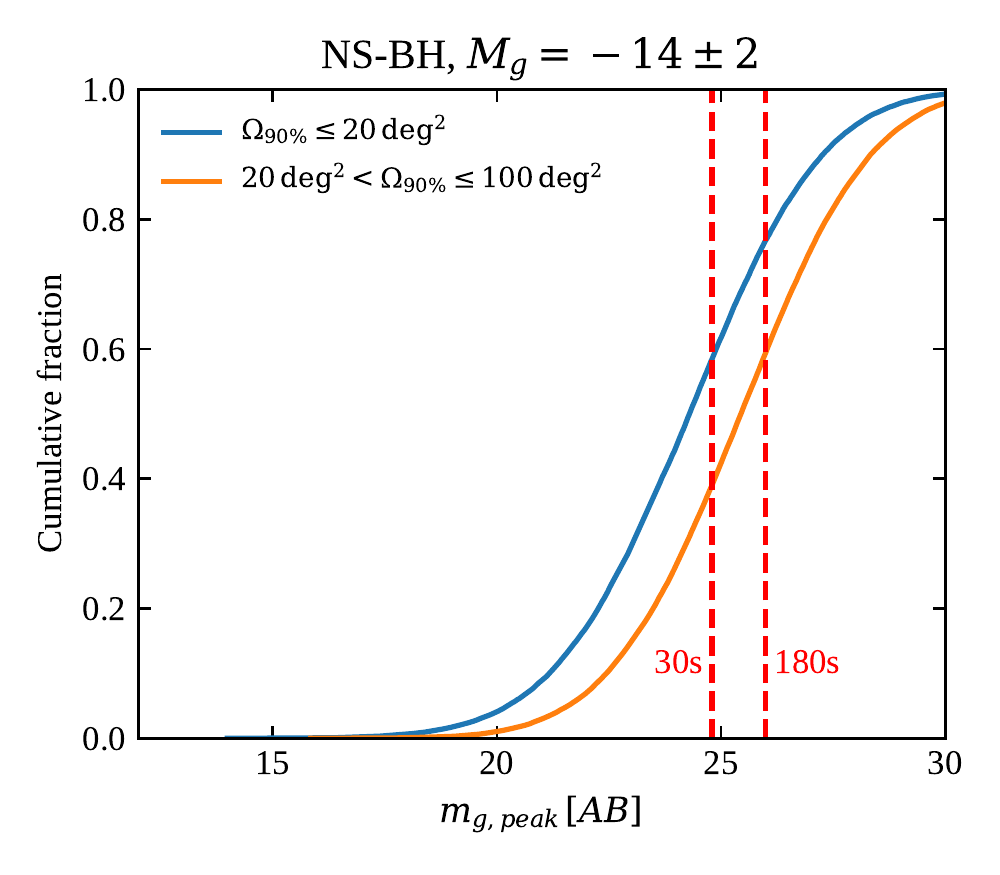}\includegraphics[width=0.48\textwidth]{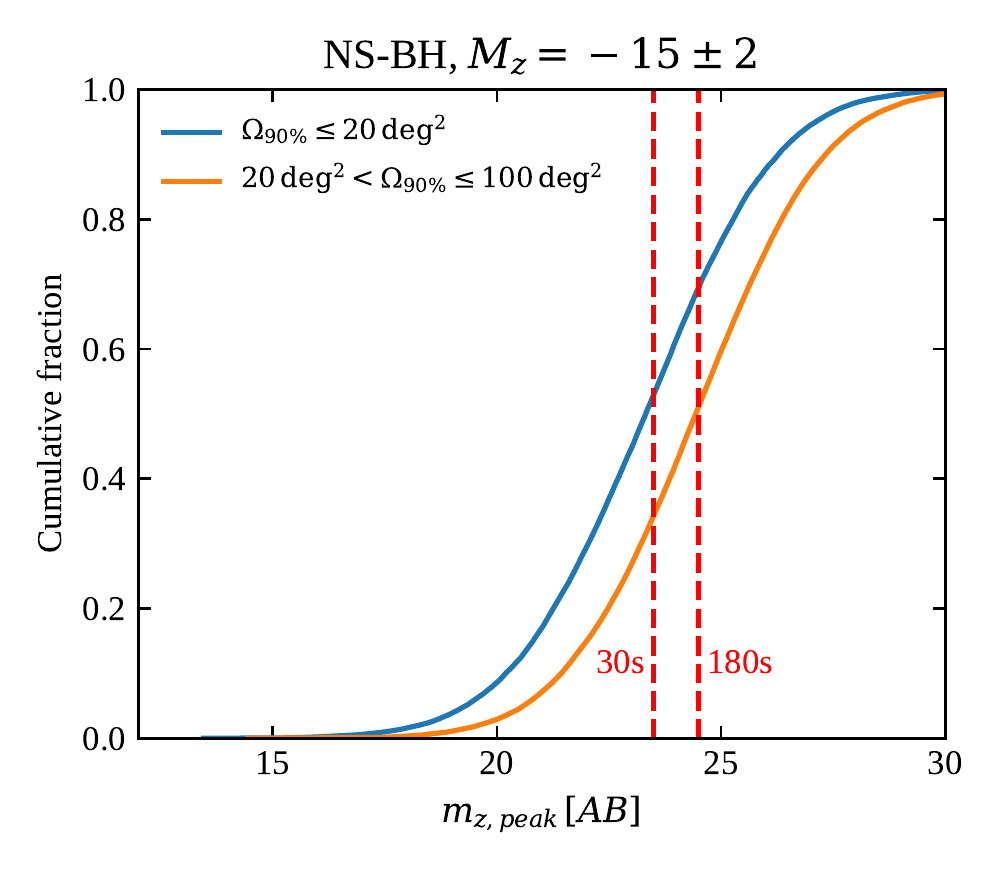}
    \caption{Cumulative apparent peak magnitude distribution of KNe associated to O5 NS--NS (top row) and NS--BH (bottom row) GW events. The left-hand panels refer to the $g$ band, while the right-hand ones are for the $z$ band. Blue lines are for tightly localized events ($\Omega_{90\%}\leq 20\,\mathrm{deg^2}$) while orange ones refer to events with $20\,\mathrm{deg^2}<\Omega_{90\%}\leq 100\,\mathrm{deg^2}$. The vertical dashed lines show our estimated 5$\sigma$ single-visit depth for 30s and 180s exposures, as annotated. }
    \label{fig:pk_mag_dist}
\end{figure*}

In order to obtain a rough estimate of our expected performance with the {\it preferred} and {\it minimal} strategy, we constructed the KN peak apparent magnitude distribution in two bands, $g$ and $z$, for $\Omega_{90\%}\leq 20\,\mathrm{deg^2}$ and
$20\,\mathrm{deg^2}<\Omega_{90\%}\leq 100\,\mathrm{deg^2}$ O5 GW-detected events separately, using the distance distributions from \cite{PeSi2021} and assuming peak absolute magnitudes $M_{g,\mathrm{peak}}=-15$\,mag and $M_{z,\mathrm{peak}}=-16$\,mag (based on AT2017gfo and our simulations,) to which we associated a Gaussian scatter with standard deviation $\sigma=1\,\mathrm{mag}$ to represent the expected intrinsic diversity of KNe \citep[][]{Gompertz2018,Ascenzi:2018mwp,Rossi2020}. The resulting distributions are shown in Figure~\ref{fig:pk_mag_dist}. 
This allows us to estimate that 97\% (96\%) of KNe associated to tightly-localized events with $\Omega_{90\%}\leq 20\,\mathrm{deg^2}$ will be detectable at peak in the $g$ band ($z$ band) with a 30\,s exposure, while the fraction decreases to 88\% (83\%) for events with $20\,\mathrm{deg^2}<\Omega_{90\%}\leq 100\,\mathrm{deg^2}$, due to the correlation between distance and  average localization accuracy. The deeper limits reached with a longer 180\,s exposure increase all these fractions to 95-100\%, but this is in part balanced by fading of the light curves after $t\gtrsim 1\,\mathrm{d}$, which is when longer exposures are performed in our proposed strategies. We therefore take the detection fractions estimated with the 30\,s exposure as reference.

The expected number  of KN detections  $N_\mathrm{det}$ is proportional to the number of ToOs and to the detection efficiency $f_\mathrm{det}$, that is
\begin{equation}
N_\mathrm{det}\sim \min\left( \frac{T_\mathrm{ToO}}{\langle T_\mathrm{single}\rangle }f_\mathrm{det}, N_\mathrm{NS-NS}\right)
\end{equation}
where $T_\mathrm{ToO}$ is the time allocated to ToOs (we are focusing here on NS--NS) and $\langle T_\mathrm{single}\rangle$ is the average time per event required to complete the observations according to the strategy. In O5, from Table~\ref{tab:detection_prospects} and accounting for the Rubin sky coverage, we expect $N_\mathrm{NS-NS}\sim 19$, of which 7 with $\Omega_{90\%}\leq 20\,\mathrm{deg^2}$ and 12 with
$20\,\mathrm{deg^2}<\Omega_{90\%}\leq 100\,\mathrm{deg^2}$. With the \textit{minimal} strategy and considering events with $\Omega_{90\%}\leq 20\,\mathrm{deg^2}$, we showed that $\langle T_\mathrm{single}\rangle \sim 1.85\,\mathrm{hours}$ and proposed a total of $T_\mathrm{ToO}=12.95\,\mathrm{hours}$. For these events, $f_\mathrm{det}=0.97$ as estimated above. Since early observations are always performed according to the strategy, $f_\mathrm{early}\sim 1$, and given the strategy characteristics, $n_\mathrm{ep}=3$ and $n_\mathrm{flt}=5$. This results in $S_\mathrm{NS-NS,min,<20\,deg^2}\sim 0.89$. For less tightly localized events ($20\,\mathrm{deg^2}<\Omega_{90\%}\leq 100\,\mathrm{deg^2}$), we have $f_\mathrm{det}=0.88$ as explained above, $f_\mathrm{early}\sim 1$ again, $n_\mathrm{ep}=3$ and $n_\mathrm{flt}=2$ (since events are only observed in 2 filters in this case). This results in $S_\mathrm{NS-NS,min,20-100\,deg^2}\sim 0.15$. Considering all events together, the overall performance of the \textit{minimal} strategy is $S_\mathrm{NS-NS,min}\sim 0.42$.

For the {\it preferred} strategy, the performance for well-localized improves thanks to the larger number of epochs, $n_\mathrm{ep}=4$, leading to $S_\mathrm{NS-NS,pref,<20\,deg^2}\sim 0.97$. For less tightly localized events, since observations are performed in more than two filters, $S_\mathrm{NS-NS,pref,20-100\,deg^2}\sim 0.22$. We note that the performance for this class of events is mainly bounded by the time investment being aimed at detecting only 3 out of 12 events, which implies $S_\mathrm{NS-NS,pref,20-100\,deg^2}\leq 0.25$ (unless more than 4 epochs are performed per event). Considering all events together, the overall performance of the \textit{preferred} strategy is $S_\mathrm{NS-NS,pref}\sim 0.5$. This number would further improve with a larger time investment for less tightly localized events. 

We adopt a performance metric with the same form for NS-BH mergers. Given the possible redder color of the associated KNe, the absence of observational constraints, and the expected wider range of luminosities \citep[e.g.][]{Barbieri2019,Barbieri2020}, we conservatively assume fainter typical peak absolute magnitudes with respect to the NS-NS case, $M_{g,\mathrm{peak}}=-14$\,mag and $M_{z,\mathrm{peak}}=-15$\,mag, and a wider scatter $\sigma=2\,\mathrm{mag}$. This, combined with the larger average distances, results in a lower detection fraction (constructed in the same way as for the NS-NS case), as shown in Fig.~\ref{fig:pk_mag_dist}. In particular, we find that 58\% (53\%) of KNe associated to tightly-localized events with $\Omega_{90\%}\leq 20\,\mathrm{deg^2}$ will be detectable at peak in the $g$ band ($z$ band) with a 30\,s exposure, while the fraction decreases to 39\% (34\%) for events with $20\,\mathrm{deg^2}<\Omega_{90\%}\leq 100\,\mathrm{deg^2}$. Adopting 180\,s exposures, as in our proposed strategy when $\Omega_{90\%}\leq 20\,\mathrm{deg^2}$, this improves to 77\% (69\%) events detectable in the $g$ band ($z$ band), and 60\% (51\%) for events with $20\,\mathrm{deg^2}<\Omega_{90\%}\leq 100\,\mathrm{deg^2}$. Assuming again $f_\mathrm{early}=1$ in all cases, these lead to $ S_\mathrm{NS-BH,min,<20\,deg^2}=0.71$, $S_\mathrm{NS-BH,min,20-100\,deg^2}=0.03$, and a combined $S_\mathrm{NS-BH,min}=0.26$ for the \textit{minimal} strategy. For the \textit{preferred} strategy, the improvements lead to $S_\mathrm{NS-BH,pref,<20\,deg^2}=0.77$ and  $S_\mathrm{NS-BH,pref,20-100\,deg^2}=0.05$, yielding a combined performance $S_\mathrm{NS-BH,pref}=0.29$. Again, the main limitation here is the time investment: if more time could be allocated to ToO's following events with a relatively coarse localisation, Rubin would be able in principle to  detect the large majority of counterparts early and to provide a multi-filter characterisation of each, revolutionising our knowledge of these sources.

For BH--BH mergers and GW events from unidentified sources for which an optical/near-IR EM counterpart has never been observed, defining the rate of success of our strategy in a similar, semi-quantitative way is not straightforward, as in this case Rubin is literally exploring the unknown.
However, we emphasize that those EM counterparts constitute a large portion of the discovery space that is made available for Rubin exploration by our ToO strategies.
Further, we expect that improvements upon these heuristics using quantitative, population-level constraints on parameters of interest, including the neutron star equation of state or the Hubble Constant may be possible in the future using these simulations \citep{DiCo2020}.

\subsection{Impact of ToOs on the LSST survey}
\label{subsec: impact}

As part of the v2.0 survey strategy simulations, we consider two simulations which include interruptions for ToO observations. For a general assessment of the impact of ToO observations on the baseline LSST survey, we consider the cases of 10 ToO events per year and 50 ToO events per year. We only attempt ToO observations for sources which fall in the main Rubin survey footprint. 

Follow-up observations are attempted in 5 filters, $g$+$r$+$i$+$y$ and $u$ or $z$ (whichever happens to be loaded depending on the moon phase). We attempt to observe in all 5 filters at a generous cadence of 0, 1, 2, 4, and 8 hours after the initial ToO alert, from which we expect an impact similar to, or greater than, the strategies described in \S\ref{sec: technical}. For 10 ToO\,yr$^{-1}$ and 50 ToO\,yr$^{-1}$, we execute 13,039 and 56,877 total visits following up ToOs, respectively.

The impact on other Rubin science cases appears to be very minimal. The number of well observed type Ia supernovae only drops to 24,800 and 24,700, in the ToO simulation, compared to 25,400 in the baseline. Other science cases such as detection of faint Near-Earth Objects (NEOs) and detection of fast microlensing events also change by only ~1\%.

\section{Discussion and Conclusion}

In this paper we presented {\it minimal} and {\it preferred} strategies for GW follow-up with Rubin Observatory.
For each type of GW detection, we outlined preferred observing cadences, exposure times, and filters as described in \S\ref{sec: technical} and summarized in Table~\ref{tab:strategies}.
ToOs with Rubin are crucial to answer the scientific questions posed in the introduction with joint EM+GW observations and will have minimal impact on the main survey (\S\ref{subsec: impact}).

Thanks to Rubin ToOs, we expect to discover counterparts to approximately 10 counterparts to NS--NS mergers and probe the existence of EM counterparts to $\sim 15$ NS--BH mergers per year during O5. The number can increase significantly if more LVK runs happen during Rubin operations. During O5, Rubin will be able to discover a larger number of counterparts via ToO observations than during the regular LSST survey, where $< 4$ KN detections per year are expected (\S\ref{subsec: high-level-description}). Un-triggered KN discovery (i.e., independent of GW or GRB detection) is important to probe EM counterparts at distances beyond the LVK horizon, helping us understand the KN luminosity function, correlations with redshift at all viewing angles, while also enabling studies of both cosmology and nuclear physics. However, Rubin ToO will provide the community with early ($\delta t<12$\,hr) and deep multi-band observations of faint KNe, and will benefit from merger time information and invaluable GW data for multi-messenger studies.

The strategies were designed to maximize the chances of discovering the EM counterpart to GW sources. As soon as the most likely counterpart is identified, a public announcement will be immediately made, allowing other facilities with large aperture but smaller FoV (e.g., Very Large Telescope, W. M. Keck Observatory, Gemini Observatory, Magellan Telescopes, ESO New Technology Telescope telescope equipped with the Son Of X-Shooter ``SOXS" spectrograph) to continue characterizing the EM transient with deep spectroscopic and photometric observations. Rubin detection of KNe will be particularly important for follow-up with space-based observatories, including James Webb Space Telescope.  Broker projects will have an important role during future GW observing runs and must commit to immediate release of data and classification whenever possible. The community will also benefit from Rubin publicly releasing the ToO follow-up strategy on each event {\it in advance} to maximize the opportunity for coordination with other ground- and space-based observatories.

The total time needed to actuate the {\it minimal} strategies is  $\sim 67$\,hr yr$^{-1}$ during O5. Assuming a GW detectors duty cycle of 0.5 during the first couple of years of Rubin operations and $\sim 8$\,hr on-sky per night, this corresponds to roughly $\sim 1.39\%$ of the LSST time budget in the first years of operations. For the {\it preferred} strategies, the total time is  $\sim 96$\,hr yr$^{-1}$, which corresponds to approximately $\sim 2\%$ of the LSST time budget. We note that these are likely upper limits to the time amount that will be required, since we expect at least some EM counterparts to be confidently identified during the first or second night of observations. Moreover, the time budget could be significantly reduced if i) an associated GRB (and ideally its afterglow) are found and are localized with $\lesssim 2$\,deg precision shortly after the GW trigger; ii) the most distant NS--NS and NS--BH mergers, which would be observable only if a very bright \citep[$M < -17.5$\,mag, see][]{Kasliwal2020} counterpart is present, are either ignored or observed only with the minimal strategies (which can be suitable to the detection of some GRB afterglows). O4 will provide further guidance on how follow-up strategies should be optimized.

The {\it preferred} strategies for NS--NS and NS--BH mergers in particular will provide a dataset that will enable modeling of the elusive blue KN component. Importantly, highly cadenced multi-filter observations on the first night and continued observations $>48$\,hr from the merger could be the only way to single out a KN candidate among the large number of supernovae and other contaminant transients found during the search \citep[see, e.g.,][]{Cowperthwaite2018contam}, which will be too faint for spectroscopic follow-up in the vast majority of cases. 

We argue that the proposed follow-up strategies, thanks to repeated multi-band observations on the first night, will enable the discovery of afterglows if a short GRB is also detected and is associated with the GW event (i.e., under favorable viewing angles). A comprehensive study of Rubin strategies to discover GRB afterglows associated with GW triggers, especially discussing the case of off-axis jets \citep[see for example][]{Ghirlanda2015, Lamb2018MNRAS, Zhu2021arXiv}, is beyond the scope of this work.

We expect that any major modification of the observing strategies proposed in this work could have a highly disruptive impact on the capability to reach Rubin multi-messenger scientific objectives. The impact of the ToO program described here on other programs is small, since observations acquired as ToOs can be used as part of other LSST surveys (\S\ref{subsec: impact}).
When a procedure for performing ToO observations with Rubin has been set for GW follow-up, a similar procedure (although with different strategies) can be applied to other special EM or multi-messenger events such as, for instance, high-energy neutrinos from astrophysical sources \citep[e.g.,][]{Stein2021NatAs}. 

Finally, plan to re-evaluate the ToO triggering criteria and observing strategies proposed here at the end of O4 and on a yearly base after the start of Rubin operations.

\section*{Acknowledgments}

We thank Lynne Jones for her work on LSST strategy simulations.  This paper was created in the nursery of the Rubin LSST  Transient and Variable Star Science Collaboration \footnote{\url{https://lsst-tvssc.github.io/}}. The authors acknowledge the support of the Vera C. Rubin Legacy Survey of Space and Time Transient and Variable Stars Science Collaboration that provided opportunities for collaboration and exchange of ideas and knowledge and of Rubin Observatory in the creation and implementation of this work.
The authors acknowledge the support of the LSST Corporation, which enabled the organization of many workshops and hackathons throughout the cadence optimization process by directing private funding to these activities.

R.~M. acknowledges support from the National Science Foundation under Grant No. AST-1909796 and AST-1944985, and by the Heising-Simons foundation.
M.~W.~C acknowledges support from the National Science Foundation with grant numbers PHY-2010970 and OAC-2117997. A.~C. acknowledges support from the NSF award AST \#1907975. S.~J.~S. acknowledges funding from STFC Grants ST/T000198/1  and ST/S006109/1. 
D.~M.\ acknowledges NSF support from grants PHY-1914448 and AST-2037297
K.~M. acknowledges support from EU H2020 ERC grant no. 758638.
A.~H. is partially supported by a Future Investigators in NASA Earth and Space Science and Technology (FINESST) award \#\,80NSSC19K1422.
M.~N. acknowledges support from the European Research Council (ERC) under the European Union’s Horizon 2020 research and innovation programme (grant agreement No.~948381) and a Fellowship from the Alan Turing Institute.
P~D'A. acknowledges support from PRIN-MIUR 2017 (grant 20179ZF5KS) and from the Italian Space Agency, contract ASI/INAF n. I/004/11/4.
E.~C.~K. and A.~G. acknowledge support from the G.R.E.A.T research environment funded by {\em Vetenskapsr\aa det}, the Swedish Research Council, under project number 2016-06012, and support from The Wenner-Gren Foundations. M.~B. acknowledges support from the Swedish Research Council (Reg. no. 2020-03330).
The UCSC team is supported in part by NASA grant NNG17PX03C, NSF grant AST-1815935, the Gordon \& Betty Moore Foundation, the Heising-Simons Foundation, and by a fellowship from the David and Lucile Packard Foundation to R.J.F.

\software{LSST metrics analysis framework \citep[MAF;][]{Jones2014SPIE}; \texttt{astropy} \citep{2013A&A...558A..33A}}; \texttt{matplotlib}; \texttt{ligo.skymap}\footnote{\url{lscsoft.docs.ligo.org/ligo.skymap}}

\bibliographystyle{aasjournal}
\begin{small}
\bibliography{references, references2, ref_WP}
\end{small}

\end{document}